\documentclass[preprint,12pt,authoryear]{elsarticle}

\journal{Engineering Applications of Artificial Intelligence}
\usepackage{amsmath}

\usepackage{amssymb}
\usepackage{comment}
\usepackage{color,soul}
\usepackage{mwe}
\usepackage{svg}
\usepackage[hidelinks]{hyperref}
\usepackage{mathtools}
\usepackage{balance}
\usepackage{multirow}

\usepackage[update,prepend]{epstopdf} 
\usepackage{fontawesome} 
\usepackage{circuitikz} 
\usepackage{tikz}
\usetikzlibrary{arrows.meta, decorations.markings}
\usepackage{cprotect} 
\usepackage[super]{nth}
\usepackage{graphicx}
\usepackage{subcaption}
\usepackage{algorithm,xpatch}
\usepackage{algpseudocode}
\usepackage{listings}
\usepackage{placeins}
\usepackage{booktabs}
\usepackage{breakcites}
\usepackage{url}

\usepackage{array}
\newcolumntype{P}[1]{>{\centering\arraybackslash}p{#1}}

\usepackage{psfrag}

\tikzstyle{block} = [draw, rectangle, 
    minimum height=3em, minimum width=5em]
\tikzstyle{sum} = [draw, circle, node distance=1cm]
\tikzstyle{input} = [coordinate]
\tikzstyle{output} = [coordinate]
\tikzstyle{tmp} = [coordinate]
\tikzstyle{pinstyle} = [pin edge={to-,thin,black}]

\xpatchcmd{\algorithmic}{\setcounter}{\algorithmicfont\setcounter}{}{}
\providecommand{\algorithmicfont}{}
\providecommand{\setalgorithmicfont}[1]{\renewcommand{\algorithmicfont}{#1}}

\newcommand{\T}{^\mathsf{T}} 
\newcommand{\R}{\mathds{R}} 

\DeclareMathOperator*{\argmin}{arg\,min}
\DeclareMathOperator*{\argmax}{arg\,max}
\usepackage{dsfont}

\begin{document}
\setalgorithmicfont{\scriptsize\sffamily}
\begin{frontmatter}

\title{Efficient safe learning for controller tuning with experimental validation}


\author[NTNU]{Marta Zagorowska}
\author[INSP]{Christopher König}
\author[ETH]{Hanlin Yu}
\author[INSP]{Efe C. Balta}
\author[ZHAW]{Alisa Rupenyan}
\author[ETH]{John Lygeros}

\address[NTNU]{Department of Engineering Cybernetics, Norwegian University of Science and Technology, Trondheim, Norway, marta.zagorowska@ntnu.no (corresponding author)}
\address[INSP]{Control and Automation Group, Inspire AG, Switzerland, efe.balta@inspire.ch} 
\address[ETH]{Automatic Control Laboratory, ETH Z\"{u}rich, Switzerland,  hanlyu@student.ethz.ch, lygeros@ethz.ch}
\address[ZHAW]{ZHAW Centre for Artificial Intelligence, ZHAW Z\"{u}rich University of Applied Sciences, Switzerland, rupn@zhaw.ch}

\begin{abstract}
Optimization-based controller tuning is challenging because it requires formulating optimization problems explicitly as functions of controller parameters. Safe learning algorithms overcome the challenge by creating surrogate models from measured data. To ensure safety, such data-driven algorithms often rely on exhaustive grid search, which is computationally inefficient. In this paper, we propose a novel approach to safe learning by formulating a series of optimization problems instead of a grid search. We also develop a method for initializing the optimization problems to guarantee feasibility while using numerical solvers. The performance of the new method is first validated in a simulated precision motion system, demonstrating improved computational efficiency, and illustrating the role of exploiting numerical solvers to reach the desired precision. Experimental validation on an industrial-grade precision motion system confirms that the proposed algorithm achieves 30\% better tracking at sub-micrometer precision as a state-of-the-art safe learning algorithm, improves the default auto-tuning solution, and reduces the computational cost seven times compared to learning algorithms based on exhaustive search.

\end{abstract}


\end{frontmatter}



\section{Introduction}
\label{sect:intro}
Optimizing controller parameters while ensuring safety is an important task in manufacturing, especially in systems where sub-micrometer precision is required. Solving such optimization problems can be challenging because of the unavailability of explicit formulations with respect to controller parameters and the computational cost of finding a solution~\citep{Advanced_Skogestad2023}. This paper builds on Bayesian optimization to find a constrained optimum of an unknown but measurable function in an efficient way.

The idea of using surrogate models for safety has been used in reliability engineering~\citep{xu2021machine}. Several algorithms using Gaussian processes to approximate unknown safety constraints have been proposed, among others by~\cite{bichon2011efficient},~\cite{fauriat2014ak}, and~\cite{azizsoltani2018adaptive}. The main focus of these algorithms was to explore the search space safely, without explicitly finding the optimum. The SafeOpt algorithm proposed by~\cite{Safe_Sui2015} and extended by~\cite{Berkenkamp_2016} is an iterative algorithm that uses Gaussian processes to learn the unknown functional form of both the objective and the constraints. It builds on the Bayesian optimization algorithm proposed by~\cite{Information_Srinivas2012}, where the upper bound of the Gaussian process corresponding to the objective function was considered to find the next iterate, without explicitly considering its safety. SafeOpt ensures the safety of chosen points in every iteration based on confidence intervals from the Gaussian processes. The choice of the new points is made by analyzing the safety of selected points from the whole search space, which requires evaluation over possibly large parameter sets~\citep{azizsoltani2018adaptive}. Since looking at the whole search space is equivalent to performing an exhaustive search, existing methods can quickly become computationally expensive, as indicated by~\cite{Safe_Fiducioso2019} and~\cite{BayesianBerkenkamp2021}.  

The version of SafeOpt from~\cite{safeopt} requires knowledge about Lipschitz constants of the underlying functions, which limited the use of SafeOpt in practical settings. A modification proposed by~\cite{Berkenkamp_2016} removed the need for knowing the Lipschitz constants and has been successfully used for controller tuning, as shown by~\cite{Performance_Khosravi2022}. A review of approaches to using Bayesian optimization to controller tuning was done by~\cite{mesbah2022fusion}. Experimental performance of Bayesian optimization in controller tuning was shown by~\cite{Controller_Fujimoto2022} where the authors initialized Gaussian processes with a simplified model of a plant. These simplifications improve the computational cost by modifying how the next iterate is found in SafeOpt. However, these simplifications still rely on a discretization of the search space, which limits their practical usage. 

A review of methods related to SafeOpt is given in~\cite{Safe_Kim2021}, where most of the reported algorithms use a discretized search space to find the optimum. The idea of merging optimization with safe learning was explored by~\cite{Constrained_Duivenvoorden2017}, where the recommended point is computed by solving auxiliary optimization problems with particle swarm methods. The method preserves the idea of SafeOpt to use confidence intervals of Gaussian processes in every iteration but redefined the way of choosing new points to make it suitable for particle swarm methods. 
Due to the heuristic nature of particle swarm methods, the approach needs adjustments to ensure good performance of the swarm. 

A grid-free version of SafeOpt using the solutions of local optimization problems to improve run time has been recently proposed by~\cite{zagorowska2022efficient}. The exhaustive search has been reformulated as a series of optimization problems to find the next recommended point. The reformulation allows avoiding heuristics while preserving the way the new points are chosen in every iteration of SafeOpt and improving the computational performance of the algorithm. The grid-free method has been shown to work in simulation~\citep{zagorowska2022efficient}. However, the influence of solving local optimization problems instead of global grid search has not yet been analyzed in detail or demonstrated in real-world experiments. 

In this work, we extend the grid-free SafeOpt from~\cite{zagorowska2022efficient} to include a systematic way of initializing the local optimization problems. The contributions of the paper are:
\begin{itemize}
    \item We reformulate SafeOpt as a series of optimization problems preserving the way new points are chosen in every iteration;
    \item We propose an initialization method for the reformulated algorithm ensuring feasibility in every iteration;
    \item We demonstrate the performance of the reformulated algorithm in a real precision motion system.
\end{itemize}
The results are compared with a benchmark algorithm for controller tuning for precision motion systems developed by~\cite{könig2021safe}, and with the industrial autotuner provided by the equipment manufacturer.

The paper is structured as follows. We first provide the necessary background knowledge on Gaussian processes and SafeOpt in Section \ref{SafeOpt}. Section \ref{sec:LocalSolvers} introduces the grid-free reformulation of SafeOpt while Section \ref{sec:Feasibility} discusses the necessary steps for using numerical solvers in grid-free SafeOpt. Section \ref{sec:SimulatioNResults} presents the performance of the new algorithm in a simulation framework, whereas Section \ref{sec:ResultsArgus} presents the results of applying grid-free and grid SafeOpt in an experiment on a real precision motion system. Finally, the conclusions and possible future work are discussed in Section \ref{sect:concl}.

\section{Background}

\label{SafeOpt}

The optimization problem is given as~\citep{Safe_Sui2015}:
\begin{subequations}
\label{eq:OptProblem}
\begin{equation}
\label{eq:ObjFormulation}
    \min_{x \in A} g_0(x), 
\end{equation}
\begin{equation}
    \label{constraints}
    \text{s.t.  } g_j(x) \leq J_{\max},\ \forall j \in \{1,2,...,J\},
\end{equation}
\end{subequations}
where $x\in A\subset\R^n$ is a vector of decision variables from a continuous search space $A$, $J_{\max}$ is the predefined constraint limit, $g_0:\R^n\rightarrow \R$ is the objective function to be minimized and $g_j:\R^n\rightarrow \R$, $j=1,\ldots,J$ constraints that must be satisfied. It is assumed that the functional form of $g_j$, $j=0,\ldots,J$ is unknown, but we can get measurements of $g_j$ that can be used to find surrogate models based on Gaussian processes.

\subsection{Gaussian process regression}
\label{sec:GPs}
Following~\cite{Berkenkamp_2016}, we use Gaussian processes to approximate $g_j$, $j=0,\ldots,J$, using measurements. We find approximations $\tilde{g}_j(x):A\rightarrow \R$ where $j=0$ corresponds to the objective function~\eqref{eq:ObjFormulation}, while $j=1,\ldots,J$ corresponds to the constraints~\eqref{constraints}. Gaussian process regression, also called kriging, assumes that the values $\tilde{g}(x_0),\tilde{g}(x_1),\ldots,\tilde{g}(x_P)$ corresponding to different $x$ are random variables, with joint Gaussian distribution for any finite $P$. The prior information about the functions $\tilde{g}_j$ is defined by known mean $\psi_j(\cdot)$ and covariance $k_j(\cdot,\cdot)$ functions:
\begin{equation}
    \tilde{g}_j(x)\sim \text{GP}(\psi_j(x),k_j(x,x)),
\end{equation}

We assume access to noisy measurements $\hat{g}_j(x)=g_j(x)+\omega$, $\omega\sim\mathcal{N}(0,\sigma^2_{\omega})$. To use Gaussian processes corresponding to $g_j$ in optimization, we need to predict the value of $\tilde{g}_j$ at an arbitrary point $\hat{x}$ using $R$ past measurement data $\mathbf{G}_j=[\hat{g}_j(x_r)]_{r=1,\ldots,R}$. Following~\cite{gaussian_processes}, the mean and variance of the prediction at a new point $\hat{x}$ are:
\begin{subequations}
\begin{equation}
    \mu_j(\hat{x}) = \psi_j(\hat{x})+ \mathbf{k}_R(\hat{x})(\mathbf{K}_R+\mathbf{I}_R\sigma^2_{\omega})^{-1} (\mathbf{G}_j-\mathbf{\Psi}_j),
    \label{eq:mean}
\end{equation}
\begin{equation}
    \sigma^2_{R,j}(\hat{x}) = k(\hat{x},\hat{x})-\mathbf{k}_R(\hat{x})(\mathbf{K}_R+\mathbf{I}_R\sigma^2_{\omega})^{-1} \mathbf{k}_R^{\T}(\hat{x}),
        \label{eq:variance}
\end{equation}
\end{subequations}
where $\mathbf{G}_j$ is a vector of $R$ observed noisy values, $\mathbf{G}_j=[\hat{g}_j]_{j=1,\ldots,R}$, $\mathbf{\Psi}_j=[\psi_j(x_r)]_{r=1,\ldots,R}$ is a vector of mean values of the past data, $j=0,\ldots,J$, the matrix $\mathbf{K}_R$ contains the covariance of past data, $k(x_a,x_b)$, $a,b=1,\ldots,R$, $\mathbf{k}_R(\hat{x})$ contains the covariance between the new point and the past data, and $\mathbf{I}_R$ denotes identity matrix of dimension $R$. 

The mean and the variance are then used to find the lower and upper confidence bounds:
\begin{subequations}
    \label{eq:Bounds}
    \begin{equation}
    l(x,j)=\mu_j(x)-\beta\sigma_{R,j}(x),
        \label{eq:LowerBound}
    \end{equation}
    \begin{equation}
    u(x,j)=\mu_j(x)+\beta\sigma_{R,j}(x),
    \label{eq:UpperBound}
    \end{equation}
\end{subequations}
where $\beta$ corresponds to the desired confidence level. 

\subsection{SafeOpt}

We follow the SafeOpt formulation from~\cite{Safe_Sui2015} with modifications proposed by~\cite{Berkenkamp_2016}.
The algorithm uses Gaussian processes as surrogates to solve the optimization problem~\eqref{eq:OptProblem}. 
A set of safe points $S_0$ that fulfill~\eqref{constraints} is required for initialization. The safe set $S_n$ at iteration $n$ is:
\begin{equation}
    \label{safe_set}
    S_n = \bigcap_{j \in \{1,2,...,J\}} \{x \in A : u_n(x,j) \leq J_{\max}\},
\end{equation}
where $u_n(x,j)$ is the upper confidence bound of the Gaussian process that models the {$j$-th} constraint at point $x$ at iteration $n$, obtained from~\eqref{eq:UpperBound}. 
The surrogate Gaussian processes are used to define the safe sets~\eqref{safe_set} after new samples are obtained in every iteration. Depending on the chosen $\beta$ in~\eqref{eq:Bounds}, computing the safe set in iteration $n$ can be used for risk assessment, quantifying the safety of points from the search space $A$~\citep{azizsoltani2018adaptive}. 

To find the next iterate, SafeOpt defines the set of potential optimizers (minimizers) $M_n$:
\begin{equation}
    \label{minimizers_set}
    M_n = \{x \in S_n : l_n(x,0) \leq \min_{x \in S_n} u_n(x,0)\},
\end{equation}
with the lower bound $l_n(x,j)$ given in~\eqref{eq:LowerBound}.
The set of points that can expand the current safe set (expanders) is defined as $E_n$~\citep{Scalable_Sukhija2022}:
    \begin{equation}
        \label{expanders_set}
        E_n = \bigcup_{j=1}^J\{x \in S_n : |\mathcal{E}_j(x)| > 0\},
    \end{equation}
where $|\mathcal{E}_j|$ is an indicator function describing if the set $\mathcal{E}_j$ is non-empty:
    \begin{equation}
        \label{expander_definition}
        \mathcal{E}_j(x) = \{x' \in A\setminus S_n : u_{n,j,(x,l_n(x,j))}(x') \leq J_{\max}\}.
    \end{equation}
\cite{Berkenkamp_2016} defines  $u_{n,j,(\overline{x},l_n(\overline{x},j))}(x')$ as the upper confidence bound of the point $x'$ if $\overline{x}$ was added to the GP with the evaluation $l_n(\overline{x},j)$. The new auxiliary training dataset contains the previous dataset created from experiments and the upper bound of the previous GP evaluated at point $\overline{x}$ from the current safe set as an artificial observation (Table \ref{tbl:AuxDataAll}, adapted from~\cite{Berkenkamp_2016}). 

\begin{table}[!tbp]
\centering
\scriptsize\sffamily
\caption{Input for the auxiliary GP for the expanders, with the artificial observation to the right of the double line, if the current safe set contains a single point $x^1$ (adapted from~\cite{Berkenkamp_2016})}
\label{tbl:AuxDataAll}
\resizebox{\columnwidth}{!}{%
\begin{tabular}{l|lllll||lllll}
Inputs  & $x^1$    & $x^1$      & $x^1$      & $\ldots$ & $x^1$      & $\overline{x}$        & $\overline{x}$        & $\overline{x}$        & $\ldots$ & $\overline{x}$        \\ \hline
Outputs & $g_0(x^1)$ & $g_1(x^1)$ & $g_2(x^1)$ & $\ldots$ & $g_J(x^1)$ & $l_n(\overline{x},0)$ & $l_n(\overline{x},1)$ & $l_n(\overline{x},2)$ & $\ldots$ & $l_n(\overline{x},J)$
\end{tabular}
}
\end{table}

Using the sets of optimizers and expanders, the SafeOpt algorithm chooses to evaluate the point $x_n$ according to:
\begin{subequations}
    \begin{equation}
        \label{berkenkamp_next_point}
        x_{n} = \underset{x \in M_n \cup E_n}{\operatorname{argmax}}\max_{j}w_n(x,j)
    \end{equation}
    \begin{equation}
        \label{std_dev}
        w_n(x,j) = u_n(x,j)-l_n(x,j)
    \end{equation}
\end{subequations}
The iterations repeat until a termination criterion is met.

\subsection{Grid-based SafeOpt}

Grid SafeOpt relies on a discretization of the search space with a ``grid'', $\mathcal{A}\subset A$. A summary of grid-based SafeOpt is shown in Algorithm \ref{alg:SafeOpt}. The sets from~\eqref{safe_set},~\eqref{minimizers_set},~\eqref{expanders_set}, which are necessary to solve~\eqref{berkenkamp_next_point}, are obtained in iteration $n$ by doing an exhaustive search over the entire grid $\mathcal{A}$. If $|\mathcal{A}|$ is large, finding the sets $\mathcal{E}_j$ in~\eqref{expander_definition} for every point in $S_n$ can be computationally expensive, because the auxiliary GP needs to be updated every time $u_{n,j,(x,l_n(x,j))}(x')$ is calculated (line 7 in Algorithm \ref{alg:SafeOpt})~\citep{Berkenkamp_2016}.

{\linespread{1.5}
\begin{algorithm}[!tbp]
\scriptsize
 \caption{Grid-based SafeOpt\label{alg:SafeOpt} following~\cite{Berkenkamp_2016}}
 \begin{algorithmic}[1]
    \Require A grid $\mathcal{A}\subset A$ with $N$ points, initial safe set $S_0=\lbrace x^0,x^1,\ldots,x^K\rbrace\subset{\mathcal{A}}$, maximal number of iterations $C$, chosen $\beta$, desired safety threshold $J_{\max}$
    \item  Set $n\leftarrow 1$, compute $F_n=\lbrace f(x^i) \rbrace_{i=1,\ldots,K}$, $G_{n,j}=\lbrace g_j(x^i) \rbrace_{i=1,\ldots,K}$ for $j=1,\ldots,J$, set $S_n\leftarrow S_0$.
    \While{$n\!\leq\! C$}
\State Using $S_n$ and $F_n$, find $\text{GP}_f$ with lower bounds $l_n(x,0)$, upper bounds $u_n(x,0)$, from~\eqref{eq:Bounds}
\State Using $S_n$ and $G_{n,j}$ find $J$ $\text{GP}_{g,j}$ with lower bounds $l_n(x,j)$, upper bounds $u_n(x,j)$, from~\eqref{eq:Bounds}
\State Find $S_n\leftarrow \bigcap_{j \in \{1,2,...,J\}} \{x \in \mathcal{A} : u_n(x,j) \leq J_{\max}\}$
\State Using~\eqref{minimizers_set} find $M_n$
\State For $j=1,\ldots,J$, for $\overline{x}\in S_n$, create $S_{n,\text{opt}}=S_n\cup \{\overline{x}\}$, $G_{n,j,\text{opt}}=G_{n,j}\cup l_n(\overline{x},j)$. 

\hspace{-0.15cm}Using $S_{n,\text{opt}}$ and $G_{n,j,\text{opt}}$ find $\text{GP}_{g,j,\text{opt}}$ with upper bounds $u_{n,j,(\overline{x},l_n(\overline{x},j))}$. 
\State For $j=1,\ldots,J$, for $x\in S_n$, $x'\in\mathcal{A}\setminus S_n$ evaluate $\mathcal{E}_j(x)$ to obtain $E_n$
\State Solve~\eqref{berkenkamp_next_point} to obtain $x_n^r$
    \State Set $n\leftarrow n+1$, $F_n\leftarrow F_{n-1}\cup\lbrace f(x_n^r) \rbrace$, $G_{n,j}\leftarrow G_{n-1}\cup\lbrace g_j(x_n^r) \rbrace$ for $j=1,\ldots,J$, $S_n\leftarrow S_{n-1}\cup \lbrace x_n^r \rbrace$
        \EndWhile
    \item \Return $x_C^r$, $f(x_C^r)$, $S_C$
    \end{algorithmic}
\end{algorithm}
}

\section{Grid-free SafeOpt}
\label{sec:LocalSolvers}
\subsection{SafeOpt as a series of optimization problems}
\label{sec:OptimizationBasedSO}
In grid-free SafeOpt from~\cite{zagorowska2022efficient}, the expanders $E_n$ and minimizers $M_n$ are found in the entire search space $A$, instead of being constrained by the grid $\mathcal{A}$. 
The search for the next iterate from~\eqref{berkenkamp_next_point} is formulated as two optimization problems:
\begin{equation}
    \label{P_1}
    P_1 : \max_{x \in M_n\subset A} \max_{j}w_n(x,j)
\end{equation}
\begin{equation}
    \label{P_2}
    P_2 : \max_{x \in E_n\subset A} \max_{j} w_n(x,j)
\end{equation}
The new value $x_n$ is obtained as the maximizing point from $E_n\cup M_n$:
    \begin{equation}
        x_n=\underset{x}{\text{argmax}}\Big\lbrace\underset{x\in E_n}{\max}\; w(x), \underset{x\in M_n}{\max}\; w(x)\Big\rbrace.
        \label{eq:MaxW}
    \end{equation}
where $w(x):=\max_{i} w_n(x,i)$. To solve the problems $P_1$ and $P_2$ using numerical solvers, we rewrite the search space of each problem, $M_n$ and $E_n$ respectively, in the form of constraints. 
The grid-free algorithm is summarised in Algorithm \ref{alg:SafeOptRef}.

{\linespread{1.5}
\begin{algorithm}[!tbp]
\scriptsize
 \caption{Reformulated SafeOpt\label{alg:SafeOptRef} following~\cite{zagorowska2022efficient}}
 \begin{algorithmic}[1]
    \Require Initial safe set $S_0=\lbrace x^0,x^1,\ldots,x^K\rbrace\subset\mathcal{A}$, desired tolerances $\epsilon_1$, $\epsilon_2$, maximal number of iterations $C$, chosen $\beta$, desired safety threshold $J_{\max}$
    \item  Set $n\leftarrow 1$, compute $F_n=\lbrace f(x^i) \rbrace_{i=1,\ldots,K}$, $G_{n,j}=\lbrace g_j(x^i) \rbrace_{i=1,\ldots,K}$ for $j=1,\ldots,J$, set $S_n\leftarrow S_0$.
    \While{$n\!\leq\! C\!\And\!\|x_n^r\!-\!x_{n-1}^r\|\!\leq\! \epsilon_1 \!\And\! \|f(x_n^r)\!-\!f(x_{n-1}^r)\|\!\leq\! \epsilon_2$}
\State Using $S_n$ and $F_n$, find $\text{GP}_f$ with lower bounds $l_n(x,0)$, upper bounds $u_n(x,0)$, 
\State Using $S_n$ and $G_{n,j}$ find $J$ $\text{GP}_{g,j}$ with lower bounds $l_n(x,j)$, upper bounds $u_n(x,j)$
\State Solve~\eqref{P_1_optimizer_condition}, obtaining $x^*_n$ and $l^*=l_n(x^*_n,0)$
\State Solve $P^j_1$ for all $j=1,\ldots,J$ from~\eqref{P_1_k}
\State Solve $P^j_2$ for all $j=1,\ldots,J$ from~\eqref{P_2_k} 
\If{$q_j(x_2^{j*},x^{'*})\geq w_n(x_2^{j*},j)$}
    \State Solve~\eqref{next_sample_efficient} and set $x_n^r\leftarrow \argmax_{\{x_1^*,x_2^*\}}\lbrace w_n(x_2^*,k_2^*),w_n(x_1^*,k_1^*) \rbrace$
\Else
    \State Set $x_n^r\leftarrow x_1^*$  
\EndIf
    \State Set $n\leftarrow n+1$, $F_n\leftarrow F_{n-1}\cup\lbrace f(x_n^r) \rbrace$, $G_{n,j}\leftarrow G_{n-1}\cup\lbrace g_j(x_n^r) \rbrace$ for $j=1,\ldots,J$, $S_n\leftarrow S_{n-1}\cup \lbrace x_n^r \rbrace$
        \EndWhile
    \item \Return $x_n^r$, $f(x_n^r)$, $S_n$
    \end{algorithmic}
\end{algorithm}
}

\subsubsection{Minimizers}
From the definition of the safe set from~\eqref{safe_set}, we obtain that:
\begin{equation}
    x\in S_n \iff x\in A \text{ and }  \forall j=1,\ldots,J\quad u_n(x,j)\leq J_{\max}.
    \label{eq:SafeSetIneq}
\end{equation}
From the definition of the minimisers~\eqref{minimizers_set} we obtain:
\begin{equation}
    x\in M_n \iff x\in S_n \text{ and } l_n(x,0)\leq l^*,
    \label{eq:MaxDef}
\end{equation}
where:
\begin{subequations} \label{eq:InternalSafe}
\begin{align}
l^* =  \min_{z}&\;u_n(z,0),\\
\text{subject to } &  u_n(z,j)\leq J_{\max},~ \forall j =1 ,\ldots,J.
\end{align} 
\end{subequations}
We note in~\eqref{P_1} that $w_n(\cdot,i)$, $w_n(\cdot,j)$ are independent from each other for $i\neq j$. Therefore, the objective function~\eqref{P_1} can be reformulated into $J$ separate problems $P_1^k$, $k=1,2,\ldots,J$. Using~\eqref{eq:SafeSetIneq} and~\eqref{eq:MaxDef}, the minimizer problem~\eqref{P_1} becomes:
\begin{subequations}
    \label{P_1_k}
    \begin{alignat}{3}
        &P_1^k:  \quad &&  \max_{x \in A} && w_n(x,k), \label{P_1_k_obj} \\
        &\quad &&\text{subject to:}\quad && u_n(x,j) \leq J_{\max} \ \forall j \in \{1,2,...J\}, \label{P_1_safety_constraint} \\
        & && &&l_n(x,0) \leq \min_{\hat{x} \in S_n} u_n(\hat{x},0). \label{P_1_optimizer_condition}
    \end{alignat}
\end{subequations}
The solution to~\eqref{P_1} is then:
\begin{equation}
    \label{P_1_solution}
    x_1^* = \underset{k \in \{1,2,...,J\}}{\operatorname{argmax}} w_n(x_1^{k*},k)
\end{equation}
where $x_1^{k*}$ is the solution to~\eqref{P_1_k}. 

\subsubsection{Expanders}

Combining~\eqref{P_2} and~\eqref{expanders_set} we get:
\begin{subequations}
\label{eq:ExpandersRef}
\begin{align}
\underset{x}{\text{max}}&\quad  \max_{i} w_n(x,i),\\
\text{subject to}    &\quad x\in S_n,\\
&\quad |\mathcal{E}(x)|> 0,
\end{align}
\end{subequations}
where $\mathcal{E}(\cdot)$ is given by~\eqref{expander_definition}. From~\eqref{expander_definition} we notice that the set $\mathcal{E}(x)$ is non-empty if there exists at least one point $x'\in A\setminus S_n$ such that the condition:
\begin{equation}
    \forall j\; u_{n,j,(x,u_n(x,j))}(x')\leq J_{\max}
\end{equation}
is satisfied. Thus, we obtain:
\begin{subequations}
\label{eq:ExpandersRefInf}
\begin{align}
\underset{x,x'}{\text{max}}&~ \max_i w_n(x,i),\\
\text{subject to}    &~ x\in S_n, \\
&~  u_{n,j,(x,u_n(x,j))}(x')\leq J_{\max},~  \forall j=1,\ldots,J,\\
&~  x'\in A\setminus S_n.
\end{align}
\end{subequations}
From the definition of the safe set~\eqref{safe_set}, we get that:
\begin{equation}
    x'\in A\setminus S_n \iff  x'\in A \text{ and } \exists k: u_n(x',k)> J_{\max}.
\end{equation}
Then we have the following equivalence:
\begin{equation}
  \exists k: u_n(x',k)> J_{\max} \iff \max_s u_n(x',s)>J_{\max}, 
  \label{eq:SafeEquivalence}
\end{equation}
from which we obtain:
\begin{subequations}
\label{eq:ExpandersRefInfOpt}
\begin{align}
\underset{x,x'}{\text{max}}&\quad \max_{i} w_n(x,i),\\
\text{subject to}    &\quad  \max_s u_n(x,s) \leq J_{\max}, \\
&\quad  \max_s u_n(x',s) > J_{\max}, \label{eq:UnSafey} \\
&\quad  \max_{s}u_{n,s,(x,u_n(x,s))}(x')\leq J_{\max}\label{eq:ExpandersSetIneq}.\\
\end{align}
\end{subequations}
The constraint~\eqref{eq:UnSafey} is feasible if $S_n\subsetneq A$. The constraint~\eqref{eq:ExpandersSetIneq} may be infeasible if the set of expanders $G_n$ is empty~\citep{Berkenkamp_2016}. To avoid infeasibility, we relax~\eqref{eq:ExpandersRefInfOpt}:
\begin{subequations}
\label{eq:ExpandersRefFeas}
\begin{align}
\underset{x,x'}{\text{max}}&\quad q(x,x'),\\
\text{subject to}    &\quad  \max_s u_n(x,s) \leq J_{\max}, \label{eq:Safey}\\
&\quad  \max_s u_n(x',s) > J_{\max} , 
\end{align}
\end{subequations}
where:
\begin{equation}
\label{eq:AllQ}
q(x,x')=\max_{i} w_n(x,i) - \sigma\max\lbrace 0,\max_s u_{n,s,(x,l_n(x,s))}(x')-J_{\max}, \rbrace
\end{equation}
and $\sigma>0$ enables trading off feasibility and optimality. The relaxation in~\eqref{eq:AllQ} has been introduced to ensure the feasibility of~\eqref{P_2} in the case the expanders do not exist while preserving the measurement-based nature of~\eqref{expander_definition} introduced by~\cite{Berkenkamp_2016}. If $\sigma$ is too small, the optimization may return an infeasible point. Conversely, choosing a large $\sigma$ puts emphasis on feasibility, at the expense of finding the optimum. 

Doing the same reformulation as in~\eqref{P_1_k}, we get the expander problem~\eqref{P_2} formulated as a series of $J$ problems:
\begin{subequations}
    \label{P_2_k}
    \begin{alignat}{3}
        &P_2^k:  \quad &&  \max_{x,x' \in A} && q_k(x,x'), \label{P_2_k_obj} \\
        &\quad &&\text{subject to:}\quad && \max_{j \in \{1,2,...,J\}} u_n(x,j) \leq J_{\max}, \label{P_2_safety_constraint} \\
        & && &&\max_{j \in \{1,2,...,J\}} u_n(x',j) > J_{\max}, \label{P_2_unsafety_constraint}
    \end{alignat}
\end{subequations}
with:
\begin{equation}
    \label{q}
    \begin{aligned}
    q_k(x,x') = w_n(x,k)  -\sigma \max\lbrace 0, \max_{j} u_{n,j,(x,l(x,j))}(x')-J_{\max}\rbrace.
    \end{aligned}
\end{equation}
If we denote the solution of~\eqref{P_2_k} $(x_2^{k*},x_2^{'k*})$, the solution to~\eqref{P_2} is defined as:
\begin{equation}
    \label{P_2_solution}
    x_2^* = \underset{k \in \{1,2,...,J\}}{\operatorname{argmax}} w_n(x_2^{k*},k).
\end{equation}
To consider the solution $x_2^*$ in~\eqref{P_2_solution} to be an expander, it is now necessary to check the gap between $q(x_2^{*},x_2^{'*})$ from~\eqref{eq:AllQ} and $\max_i w_n(x_2^{*},i)$ from~\eqref{eq:ExpandersRefInfOpt}. If, for every $j$, $x_2^*$ and $x'^*$ fulfill:
\begin{equation}
    \label{expander_constraint}
    q_j(x_2^*,x'^*) \geq w(x_2^*,j),
\end{equation}
then the algorithm chooses $x_n$ between $x_1^*$ and $x_2^*$ by solving:
\begin{equation}
    \label{next_sample_efficient}
    x_n = \underset{\{x_1^*,x_2^*\}}{\operatorname{argmax}} \{w_n(x_1^*,k_1^*),w_n(x_2^*,k_2^*)\}.
\end{equation}
If~\eqref{expander_constraint} is not satisfied, then $u_{n,j,(x,l(x,j))}(x') > J_{\max}$ and $x_n=x_1^*$. 

The problems~\eqref{P_1_k},~\eqref{P_2_k} use the same definitions of the optimizers and the expanders as~\cite{Berkenkamp_2016}, thus preserving the safety properties of SafeOpt~\citep{BayesianBerkenkamp2021}. 

\section{Embedding numerical solvers in SafeOpt}
\label{sec:Feasibility}

\subsection{Initialization}
 
To facilitate the solution of~\eqref{P_1} and~\eqref{P_2} with local solvers, we first propose an initialization method. Let us recall that we assume some initially feasible safe set $S_0$. Optimization problems~\eqref{P_1_optimizer_condition} and subsequently~\eqref{P_1_k} (lines 5 and 6 in Algorithm \ref{alg:SafeOptRef}) are feasible for all $x\in S_n$, $n\geq 0$. Thus, starting the local solver with an initial guess $\hat{x}\in S_n$ ensures that a feasible solution exists. We choose to start the optimization problems at the current best solution, $\hat{x}=\argmin F_{n}$ where $F_n=\lbrace f(x^i) \rbrace_{i=1,\ldots,K}$ contains the values of the objective function evaluated at the samples $x^i$ obtained until iteration $n$ (line 1 in Algorithm \ref{alg:SafeOptRef}).

To overcome the limitations of using a local solver, we use the definition of the safe set~\eqref{safe_set} to find an initial guess for the expander search~\eqref{P_2_k}. Starting a local solver from infeasible points can lead to only finding points that do not fulfill~\eqref{P_2_unsafety_constraint} in~\eqref{P_2_k}. Following~\cite{Constrained_Duivenvoorden2017} and~\cite{könig2021safe}, we look for expanders on the boundary of the current safe set. 

The proposed procedure to generate a feasible starting point for the search is summarised in Algorithm \ref{alg:InitiChoice}. We sample $m$ points $x_i$ from $S_n$ (line 1) and $l$ points $x_j'$ from $A\setminus S_n$ (line 2) in every iteration. Choosing a combination of safe and unsafe points as a starting point gives feasibility of~\eqref{P_2}. To find points at the boundary of the current safe set, we look at the Euclidean distance between points chosen from $S_n$ and $A\setminus S_n$ (line 3). The points in the safe set and in the unsafe set are paired up according to the minimal Euclidean distance between them, i.e. each of $m$ safe points is paired up with the unsafe point closest to it (lines 4-7). The procedure returns a set of $m$ pairs, $\lbrace\hat{x}_i\rbrace_{i=1,\ldots,m}$ where $\hat{x}_i= [x_i, x'_{j^*}]$.

Algorithm \ref{alg:InitiChoice} provides a set of feasible points $\lbrace\hat{x}_i\rbrace_{i=1,\ldots,m}$. We choose one point from $\lbrace\hat{x}_i\rbrace_{i=1,\ldots,m}$ as a starting point to solve~\eqref{P_2_k}. We also note that multiple initial guesses can be considered in parallel.

{ \linespread{1.5}
\begin{algorithm}[!tbp]
    \caption{Generating starting points for~\eqref{P_2_k}\label{alg:InitiChoice}}
    \begin{algorithmic}[1]
    \item Generate $m$ safe points $x_i$ from $S_n$, $i \in \{1,2,...,m\}$
    \item Generate $l$ unsafe points $x_j'$ from $A\setminus S_n$, $ j \in \{1,2,...,l\}$
    \item Calculate the Euclidean distance between each safe and unsafe point and store them in a matrix $M \in \mathbb{R}^{m \times l}$
    \For{$i \in \{1,2,...,m\}$}
        \State Find $j^* \gets \underset{j \in \{1,2,...,J\}}{\operatorname{argmin}} M_{ij}$
        \State Create potential starting points $\hat{x}_i \gets [x_i, x'_{j^*}]$
    \EndFor
    \item \Return $\lbrace\hat{x}_i\rbrace_{i=1,\ldots,m}$
    \end{algorithmic}
\end{algorithm}
}

\subsection{Choice of solver}
The proposed reformulation is independent of the chosen optimization solver. We focus on derivative-free methods to preserve the derivative-free character of Bayesian optimization.
\cite{zagorowska2022efficient} explored the flexibility provided by using pattern search methods to introduce new stopping criteria for the reformulated SafeOpt algorithm. Pattern search methods belong to the group of direct search optimization methods and rely on evaluating a number of candidate points around a selected point, which are chosen from a \emph{mesh} following a given \emph{pattern}. The mesh can be intuitively understood as local discretization of adjustable size around the current point. If the size of the mesh falls below a given threshold, the algorithm stops.~\cite{Derivative_Audet2017} provide an in-depth description of the algorithm and its convergence properties. 

\section{Controller tuning with unknown constraints}
\label{sec:SimulatioNResults}

\subsection{Simulation setup}
\begin{figure}[!tbp]
    \centering
\scalebox{0.9}{
\scriptsize
    \begin{tikzpicture}[auto, node distance=3cm]
    \node [input, name=input] {};
    \node [sum, right of=input] (sum) {};
    \node [block, right of=sum,text width=2cm, align=center,node distance =1.5cm] (p_controller) {\textsf{\scriptsize{$C_p(s)$}}};
    \node [sum, right of=p_controller,node distance =1.5cm] (sum2) {};
    \node [block, right of=sum2, text width=2cm, align=center,node distance =1.5cm] (v_controller) {\textsf{\scriptsize{$C_s(s)$}}};
    \node[block, right of=v_controller, text width=2cm, align=center](system){\scriptsize\textsf{$G(s)$}};
    \node[block, right of=system,text width=2cm, align=center](trans){\scriptsize\textsf{$1/s$}};
    \node [input, above of=p_controller,name=tmpIn, node distance =1cm]{};
    \draw[->,>=Stealth] (tmpIn) -| node [pos=0.95]{} (sum2);
    \draw[-] (tmpIn) -| node [pos=0.1]{$S_s$} (sum2);
    \draw[->,>=Stealth] (input) --node {\scriptsize$P_s$} (sum);
    \draw[->,>=Stealth] (sum) --node {} (p_controller);
    \draw[->,>=Stealth] (p_controller) --node {} (sum2);
    \draw[->,>=Stealth] (sum2) --node {} (v_controller);
    \draw[->,>=Stealth] (v_controller) --node {} (system);
    \draw[->,>=Stealth] (system) --node {} (trans);
    \node[output, right of=trans, node distance=2cm](output_p){};
    \draw[->,>=Stealth] (trans) -- node {}(output_p);
    \node[tmp, right of=system, node distance=1.5cm] (tmpS){};
    \node[tmp, below of=tmpS, node distance=1cm] (tmpS2){};
    \draw (tmpS) to[short, -*] (tmpS);
    \draw[-] (tmpS) -- node[pos=0.95]{$S$} (tmpS2);
    \draw[->,>=Stealth] (tmpS2) -| node[pos=0.95]{$-$} (sum2);

    \node[tmp, right of=trans, node distance=1.5cm] (tmpP){};
    \node[tmp, below of=tmpP, node distance=1.5cm] (tmpP2){};
    \draw (tmpP) to[short, -*] (tmpP);
    \draw[-] (tmpP) -- node[pos=0.95]{$P$} (tmpP2);
    \draw[->,>=Stealth] (tmpP2) -| node[pos=0.95]{$-$} (sum);

    \end{tikzpicture}
    }
    \caption{Block diagram of a ball–screw drive with transfer function $G(s)$. The objective is to follow the position set point $P_s$ ensured by a proportional controller $C_p(s)$ in cascade with a speed controller $C_s(s)$ (adapted from \cite{zagorowska2022efficient})}
    \label{fig:BlockDiagram}
\end{figure}
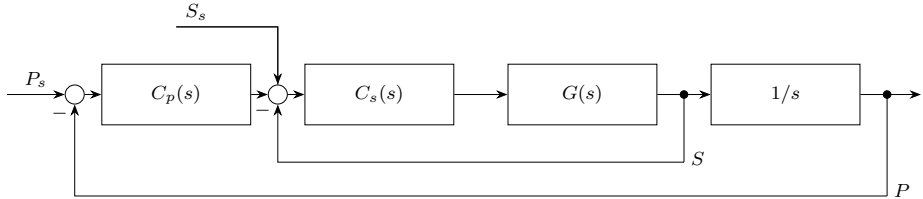

To test the proposed reformulation of SafeOpt, we first show results from a numerical simulation of tuning a cascade PID controller for the ball-screw drive from~\cite{CascadeKhosravi2020} and~\cite{zagorowska2022efficient} (Fig. \ref{fig:BlockDiagram}). The objective is to find a parameter $K_p$ for the position controller $C_p(s)$ and the parameters $K_v$ and $K_{vi}$ for the speed control $C_s(s)$ to minimize weighted average in the position error $P-P_s$ and the speed error $S$:
\begin{equation}
    J := \gamma_1\|P-P_s\|_1+ \|S\|_{\infty}.
    \label{eq:RefObjPID}
\end{equation}
To put emphasis on tracking the desired position setpoint $P_s$ in this example, we set $\gamma_1=1000$ in accordance with the magnitude of the measured signals.

To emulate human-driven PID tuning based on visual assessment of responses of the system~\citep[Ch. 4.4]{Advanced_Astroem2006}, we measure stability as the slope $p_1$ of the peaks of the response of the system, with positive values indicating instability. The constraint was formulated as:
\begin{equation}
    h(K_p,K_v,K_{vi}):=\gamma_2(p_1-\sigma)\leq 0,
    \label{eq:RefCstrPID}
\end{equation}
where $\gamma_2=100$ for scaling and $\sigma=0.005$ was chosen to ensure that a system with no peaks, i.e. $p_1=0$, yields a value inside the feasible set. 

Setting $x:=[K_p,K_v,K_{vi}]^{\T}$, we obtain the problem structure of~\eqref{eq:OptProblem}. The search space $\mathcal{A}=[0,110]\times [0,50]^2$ was chosen so that it contains unstable values. The initial safe set contains four points found using the simulation. Table \ref{tbl:InitialSafeSet} shows the values of the objective~\eqref{eq:RefObjPID} and the constraint~\eqref{eq:RefCstrPID} for the initial safe set $S_0$ and $J_{\max}=0$. For the simulations, we used squared exponential kernels with hyperparameters obtained using \texttt{fitrgp} in Matlab~\citep{zagorowska2022efficient}.

\begin{table}[!tbp]
\caption{The initial safe set $S_0$, and an unsafe point, with the corresponding value of the objective function and the constraint (adapted from \cite{zagorowska2022efficient})}
\label{tbl:InitialSafeSet}
\centering
\scriptsize\sffamily
\begin{tabular}{@{}lccccc@{}}
\toprule
             & $K_p$ & $K_v$ & $K_{vi}$ & Obj.~\eqref{eq:RefObjPID} &  Cstr. \eqref{eq:RefCstrPID}\\ \cmidrule{1-6}
Unsafe point   & 30    & 0     & 5        & 388      & 4.6        \\
Safe point I   & 10    & 0     & 5        & 241      & -4.8       \\
Safe point II  & 20    & 0.4  & 50       & 20       & -5.4       \\
Safe point III & 42    & 0.3   & 12       & 39       & -5.8       \\
Safe point IV  & 90    & 0.5   & 1        & 26       & -0.005         \\ \bottomrule
\end{tabular}
\end{table}

\begin{figure}
\centering
    \psfrag{0}[][]{\tiny\textsf{0}}
    \psfrag{2}[][]{\tiny\textsf{2}}
    \psfrag{4}[][]{\tiny\textsf{4}}
    \psfrag{6}[][]{\tiny\textsf{6}}
    \psfrag{8}[][]{\tiny\textsf{8}}
    \psfrag{10}[][]{\tiny\textsf{10}}
    \psfrag{12}[][]{\tiny\textsf{12}}

    \psfrag{0.5}[][]{\tiny\textsf{0.5}}
    \psfrag{1}[][]{\tiny\textsf{1}}
    \psfrag{1.5}[][]{\tiny\textsf{1.5}}
    \psfrag{-0.5}[][]{\tiny\textsf{-0.5}}
    \psfrag{-1}[][]{\tiny\textsf{-1}}
    \psfrag{-1.5}[][]{\tiny\textsf{-1.5}}

    \psfrag{0.2}[][]{\tiny\textsf{0.2}}
    \psfrag{0.4}[][]{\tiny\textsf{0.4}}
    \psfrag{0.6}[][]{\tiny\textsf{0.6}}
    \psfrag{0.8}[][]{\tiny\textsf{0.8}}
    \psfrag{-0.2}[][]{\tiny\textsf{-0.2}}

    \psfrag{1.45}[][]{\tiny\textsf{1.45}}
    \psfrag{1.4}[][]{\tiny\textsf{1.4}}
    \psfrag{1.35}[][]{\tiny\textsf{1.35}}
    \psfrag{1.3}[][]{\tiny\textsf{1.3}}
    \psfrag{1.25}[][]{\tiny\textsf{1.25}}
    \psfrag{5.2}[][]{\tiny\textsf{5.2}}
    \psfrag{5.4}[][]{\tiny\textsf{5.4}}
    \psfrag{5.6}[][]{\tiny\textsf{5.6}}
    \psfrag{7.8}[][]{\tiny\textsf{7.8}}
    \psfrag{7.9}[][]{\tiny\textsf{7.9}}

    \psfrag{-0.02}[][]{\tiny\textsf{-0.02}}
    \psfrag{0.02}[][]{\tiny\textsf{0.02}}
    \psfrag{0.04}[][]{\tiny\textsf{0.04}}

    \psfrag{Speed [cm s}[][]{\scriptsize\textsf{Speed $\Big[$cm s}}
    \psfrag{]}[][]{\scriptsize\textsf{$\Big]$}}
    \psfrag{Time [s]}[][]{\scriptsize\textsf{Time [s]}}
    \psfrag{Position [m]}[][]{\scriptsize\textsf{Position [m]}}

    \psfrag{RA 10}[][]{\scriptsize\textsf{RA 10}}
    \psfrag{RA 1}[][]{\hspace{0.05cm}\scriptsize\textsf{RA 1}}
    \psfrag{N=10}[][]{\scriptsize\textsf{$N=$10}}
    \psfrag{N=1000}[][]{\scriptsize\textsf{$N=$1000}}
    \psfrag{N=2500}[][]{\scriptsize\textsf{$N=$2500}}
          \begin{subfigure}[b]{0.8\textwidth}
         \centering
         \includegraphics[width=\textwidth]{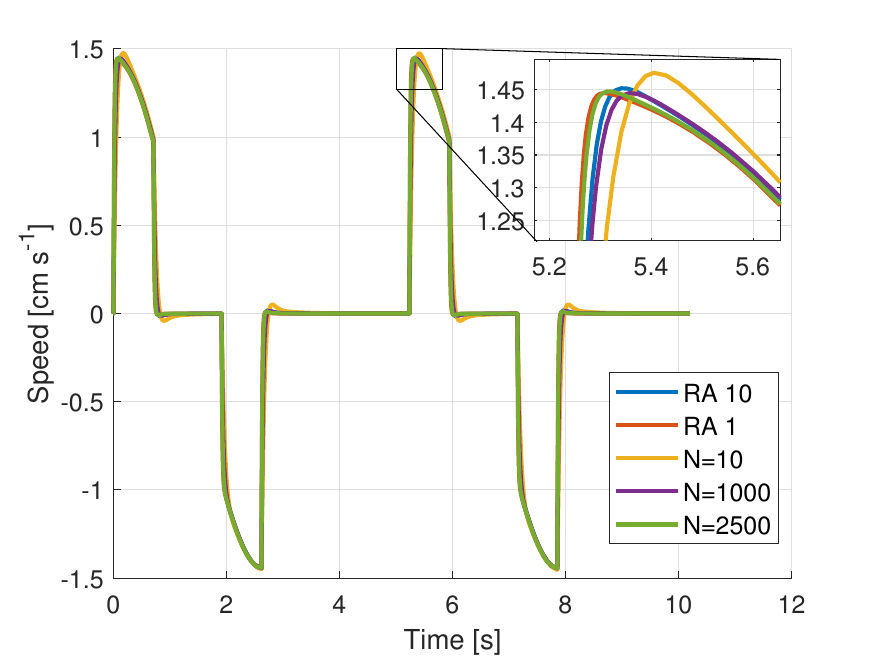}
         \caption{Speed }
         \label{fig:SpeedResSin1}
     \end{subfigure}
     ~
     \begin{subfigure}[b]{0.8\textwidth}
         \centering
         \includegraphics[width=\textwidth]{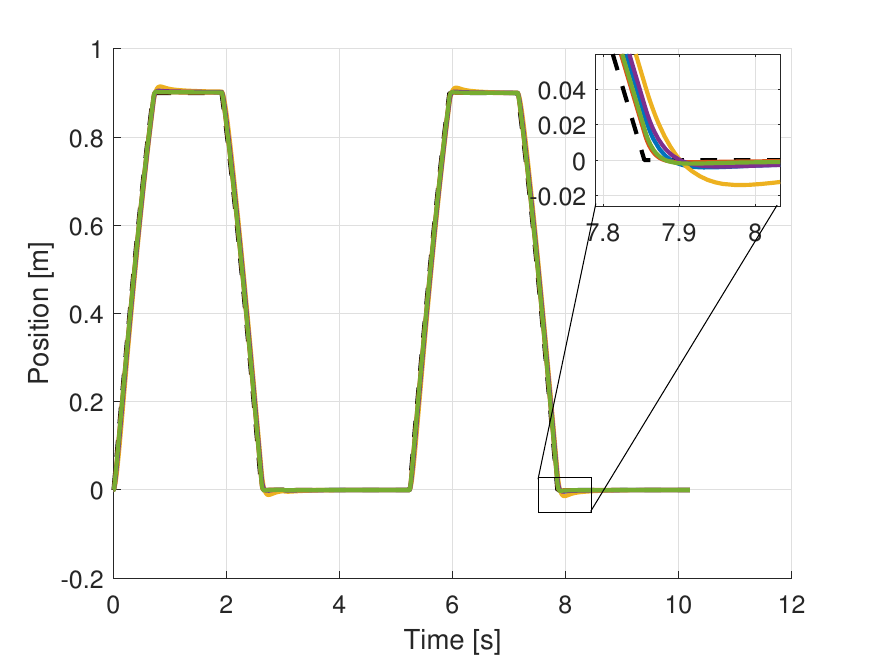}
         \caption{Position }
         \label{fig:PositionResSin1}
     \end{subfigure}
        \caption{Speed and position for the truncated sinusoidal position set point}
        \label{fig:ResultsSinPID}
        
\end{figure}

\subsection{Performance of grid-free SafeOpt}
All tests in this section were performed in Windows 10, using Matlab 2022a on a laptop with an AMD Ryzen 7 PRO 5850U, 8 cores, with 32\,GB of RAM. The model of the controller from Fig. \ref{fig:BlockDiagram} was developed in Simulink version 2022a, update 5.

Following the performance comparison from~\cite{zagorowska2022efficient}, the parameters of the reformulated algorithm were chosen as $\epsilon_1=\epsilon_2=0.1$, with a mesh tolerance of 0.01 and the initial mesh sizes of 10 (RA 10) and one (RA 1). A comparison with grid-based SafeOpt for three grids with $N\in\{10,1000,2500\}$ is shown in Fig. \ref{fig:ResultsSinPID}. The position trajectory $P_s$ was a sinusoidal function, truncated at 0.9 and zero (dashed black in Fig. \ref{fig:PositionResSin1}). The best results in terms of both the objective and the computational time (value 3.4 obtained for 17 s) were obtained for the grid-free version of SafeOpt with the initial mesh size equal to one (orange). The default SafeOpt was second best in terms of the objective (green line), at the expense of the computational time (value of 4.3, obtained in 56 s).

\subsection{Impact of initialization}
All tests in this section were performed in Windows 11 Pro, using Matlab 2022a on a laptop with an 11th Gen Intel(R) Core(TM) i7-1165G7 processor, with 32\,GB of RAM. The model of the controller from Fig. \ref{fig:BlockDiagram} was developed in Simulink version 2022a, update 5.

To further analyse the performance of the reformulation from Section \ref{sec:OptimizationBasedSO}, we implemented Algorithm \ref{alg:InitiChoice} in the simulation setup. The performance was evaluated from the perspective of the timings for solving~\eqref{P_2}, averaged over ten runs, and the resulting safe set after the algorithm has converged. The stopping criteria for SafeOpt were $\epsilon_1=\epsilon_2=0.1$, and for pattern search constraint tolerance $\varepsilon=0.01$, and minimum mesh size $\delta^0=1$. We used two kinds of sampling from the search space $A$ in Algorithm \ref{alg:InitiChoice}: Latin hypercube sampling from~\cite{osti_7091452} (using \texttt{lhsdesign}) and random sampling (using \texttt{rand}). The $m$ safe points were obtained by evaluating~\eqref{safe_set} for the $m_0$ points. The number of unsafe points was then $l=m_0-m$. We chose $m_0\in\lbrace 100,300,500,700 \rbrace$ (rows 2-5 in Table \ref{tbl:resultsAlgorithm}) and paired it with a fixed unstable point from Table \ref{tbl:InitialSafeSet} used as a guess for $x'$ in the expander search (top row in Table \ref{tbl:InitialSafeSet}).

\subsubsection{Impact on timing}

\begin{table}[!tbp]
\centering
\scriptsize\sffamily
\caption{Timings of implementation of Algorithm \ref{alg:InitiChoice} to generate an initial guess for expander search (first four rows), together with a fixed unstable point from Table \ref{tbl:InitialSafeSet} used as a guess for $x'$ (last row), using Latin hypercube sampling ($LH$) and random sampling ($R$)}
\label{tbl:resultsAlgorithm}
\resizebox{\columnwidth}{!}{%
\begin{tabular}{@{}ccccc@{}}
\toprule
$m_0$ & Resulting safe set [\%] & Time [s] for Algorithm \ref{alg:InitiChoice} & Time [s]  for solving~\eqref{P_2} & Overall time [s] \\ \midrule
100   & 31 $_{LH}|_{R}$           28           & 0.00087 $_{LH}|_{R}$  0.00086                                 & 0.04  $_{LH}|_{R}$ 0.05                            & 2.9 $_{LH}|_{R}$ 3.7             \\
300   & 5 $_{LH}|_{R}$ 76                      & 0.0009 $_{LH}|_{R}$ 0.001                                 & 0.04  $_{LH}|_{R}$ 0.06                            & 2.4   $_{LH}|_{R}$ 2           \\
500   & 15  $_{LH}|_{R}$ 35                     & 0.0014  $_{LH}|_{R}$ 0.0011                                    & 0.03 $_{LH}|_{R}$ 0.03                             & 2.5 $_{LH}|_{R}$ 2.4             \\
700   & 0.3 $_{LH}|_{R}$  5                  & 0.0012  $_{LH}|_{R}$   0.0014                                 & 0.03      $_{LH}|_{R}$   0.05                     & 3   $_{LH}|_{R}$ 2.73             \\
-     & 28                      & n/a                                         & 10                                & 27               \\ \bottomrule
\end{tabular}%
}
\end{table}

Algorithm \ref{alg:InitiChoice} was implemented by first sampling $m_0$ points from the search space using Latin hypercube sampling. The results of running the Algorithm \ref{alg:SafeOptRef} with Algorithm \ref{alg:InitiChoice} used for starting the expander search are shown in Tables \ref{tbl:resultsAlgorithm} and \ref{tbl:resultsAlgorithmValues}. In all the cases, the reformulated SafeOpt needed $T=4$ iterations, and all iterations have found expanders. The results from all the cases were also close, with the objective function 5.69, the parameters $K_p=62$, $K_v=32.36$, $K_{vi}=50$ if Algorithm \ref{alg:InitiChoice} was used and 5.64 the parameters $K_p=62.25$, $K_v=32.38$, $K_{vi}=50$ if the starting point was fixed.

The impact of using Algorithm \ref{alg:InitiChoice} is primarily visible in the timing for solving the expander search~\eqref{P_2}, which is then propagated to the overall time (columns 3 and 4 in Table \ref{tbl:resultsAlgorithm}, respectively). Starting the expander search from a fixed initial point required nine times as long as starting on the boundary of the current safe set. The initialization based on generating and evaluating $m_0$ points in Algorithm \ref{alg:InitiChoice} is thus faster than starting the optimization problem from a fixed point. Moreover, the overall time for SafeOpt using Algorithm \ref{alg:InitiChoice} remains similar regardless of $m_0$.

\begin{table}[!tbp]
\centering
\scriptsize\sffamily
\caption{Results of implementation of Algorithm \ref{alg:InitiChoice} to generate an initial guess for expander search (first four rows), together with a fixed unstable point from Table \ref{tbl:InitialSafeSet} used as a guess for $x'$ (last row), using Latin hypercube sampling ($LH$) and random sampling ($R$)}
\label{tbl:resultsAlgorithmValues}
\resizebox{\columnwidth}{!}{%
\begin{tabular}{@{}ccccccc@{}}
\toprule
$m_0$ & $T$ & $\#$ iterations with expanders & $K_p^*$            & $K_v^*$                  & $K_{vi}^*$         & Objective                 \\ \midrule
100   & 4 $_{LH}|_{R}$ 4     & 4 $_{LH}|_{R}$ 4    & 62 $_{LH}|_{R}$ 62 & 32.36 $_{LH}|_{R}$ 32.36 & 50 $_{LH}|_{R}$ 50 & 5.7 $_{LH}|_{R}$ 5.7  \\
300   & 4 $_{LH}|_{R}$ 2     & 4 $_{LH}|_{R}$ 2    & 62 $_{LH}|_{R}$ 20 & 32.36 $_{LH}|_{R}$ 0.36  & 50 $_{LH}|_{R}$ 50 & 5.7 $_{LH}|_{R}$ 20 \\
500   & 4 $_{LH}|_{R}$ 4     & 4 $_{LH}|_{R}$ 4    & 62 $_{LH}|_{R}$ 62 & 32.36 $_{LH}|_{R}$ 32.36 & 50 $_{LH}|_{R}$ 50 & 5.7 $_{LH}|_{R}$ 5.7  \\
700   & 4 $_{LH}|_{R}$ 4     & 4 $_{LH}|_{R}$ 3    & 62 $_{LH}|_{R}$ 62 & 32.36 $_{LH}|_{R}$ 32.36 & 50 $_{LH}|_{R}$ 50 & 5.7 $_{LH}|_{R}$ 5.7  \\
-     & 4                    & 4                   & 62.25              & 32.38                    & 50                 & 5.6                     \\ \bottomrule
\end{tabular}%
}
\end{table}

The impact of the starting guesses in solving~\eqref{P_2} is visible if random sampling was used in Algorithm \ref{alg:InitiChoice}. Choosing $m_0=300$ by using random sampling resulted in points that made finding an improved solution impossible. In particular, in the first iteration of SafeOpt, the recommended value from~\eqref{next_sample_efficient} was chosen as a minimizer from the initial safe set in Table \ref{tbl:InitialSafeSet}. As we assumed no noise in the simulation, the new measurement obtained from applying the recommended value was identical to a measurement corresponding to the safe point. Thus, it triggered the stopping criterion defined by $\epsilon_2$ and the algorithm stopped. The resulting controller parameters are selected as one of the initial safe points (second row in Table \ref{tbl:resultsAlgorithmValues}) and the corresponding value of the minimized objective function is large (last column in Table \ref{tbl:resultsAlgorithmValues}). Even though the case with no noise is rarely encountered in practice, a possible remedy is to use the number of iterations as a stopping criterion at the expense of increased time. 

\subsubsection{Impact on safe sets}

The performance of using Gaussian processes as surrogates to quantify safety depends on the sampled points used in computation of~\eqref{eq:mean} and~\eqref{eq:variance}. From~\eqref{next_sample_efficient}, we see that the sets of expanders and optimizers define the sample in iteration $T-1$, thus affecting the safe set $S_T$ in iteration $T$. We use the safe set $S_T$ to quantify the impact of Algorithm \ref{alg:InitiChoice} on grid-free SafeOpt by evaluating~\eqref{safe_set} for 500000 points sampled from the entire search space using Latin hypercube sampling from~\cite{osti_7091452}. 

\begin{figure}[!tbp]
    \centering
        \psfrag{vi}[][]{\scriptsize\textsf{$vi$}}
    \psfrag{p}[][]{\scriptsize\textsf{$p$}}
    \psfrag{v}[][]{\scriptsize\textsf{$v$}}
    \psfrag{K}[][]{\hspace*{-0.2cm}\scriptsize{\textsf{$K$}}}
    \psfrag{\%/100}[][]{\scriptsize{\textsf{$\%\cdot 100$}}}
    \psfrag{0}[][]{\tiny{\textsf{0}}}
    \psfrag{0.1}[][]{\tiny{\textsf{0.1}}}
    \psfrag{0.2}[][]{\tiny{\textsf{0.2}}}
    \psfrag{0.3}[][]{\tiny{\textsf{0.3}}}
    \psfrag{0.4}[][]{\tiny{\textsf{0.4}}}
    \psfrag{0.5}[][]{\tiny{\textsf{0.5}}}
    \psfrag{0.6}[][]{\tiny{\textsf{0.6}}}
    \psfrag{0.7}[][]{\tiny{\textsf{0.7}}}
    \psfrag{0.75}[][]{\tiny{\textsf{0.75}}}
    \psfrag{0.8}[][]{\tiny{\textsf{0.8}}}
    \psfrag{0.85}[][]{\tiny{\textsf{0.85}}}
    \psfrag{0.9}[][]{\tiny{\textsf{0.9}}}
    \psfrag{1}[][]{\tiny{\textsf{1}}}
    \psfrag{-3}[][]{\tiny{\textsf{-3}}}
    \psfrag{-4}[][]{\tiny{\textsf{-4}}}

    \psfrag{0.01}[][]{\tiny\textsf{0.01}}
    \psfrag{0.02}[][]{\tiny\textsf{0.02}}
    \psfrag{0.04}[][]{\tiny\textsf{0.04}}
    \psfrag{0.78}[][]{\tiny\textsf{0.78}}
    \psfrag{0.82}[][]{\tiny\textsf{0.82}}
    \psfrag{0.84}[][]{\tiny\textsf{0.84}}
    \psfrag{5}[][]{\tiny{\textsf{5}}}
    \psfrag{10}[][]{\tiny{\textsf{10}}}
    \psfrag{15}[][]{\tiny{\textsf{15}}}

        \includegraphics[width=0.8\textwidth]{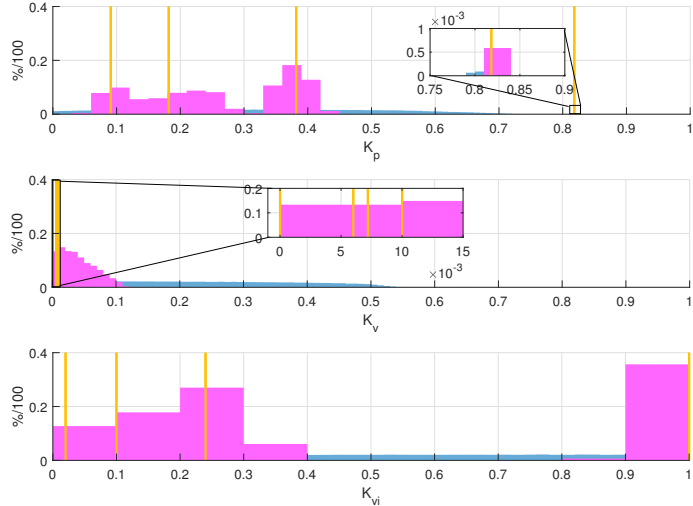}
        \caption[Iteration time comparison]
        {(Color online) The final safe set obtained from SafeOpt if Algorithm \ref{alg:InitiChoice} was used with $m_0=100$ (blue) and $m_0=700$ (magenta) points. The initial safe set $S_0$ is marked with yellow}
        \label{fig:SafeSetsComparison100700}
\end{figure}

The impact of $m_0$ in Algorithm \ref{alg:InitiChoice} is visible in the resulting safe set (second column in Table \ref{tbl:resultsAlgorithm}). If $m_0=100$, the final set of parameters considered safe covered a third of the entire search space. Conversely, choosing $m_0=700$ led to a safe space of 0.3\% (magenta in Fig. \ref{fig:SafeSetsComparison100700}). Figure \ref{fig:SafeSetsComparison100700} shows that for 100 points,  the region of the search space covered for 100 points is large (blue in Fig. \ref{fig:SafeSetsComparison100700}). This is because the initial guess for the expander search was farther from the initial safe set. Conversely, the final safe set obtained for 700 points (magenta) is clustered around the initial safe set (yellow circles). The clustering is especially prominent in the case of $K_v$ (middle plot). The clustering is due to the choice of the value of $K_v$ in the initial set close to zero (third column in Table \ref{tbl:InitialSafeSet}). The expanders remained close to zero because the local search was started close to the initial safe set thanks to 700 points. 

\begin{figure}[!tbp]
    \centering
    \psfrag{vi}[][]{\scriptsize\textsf{$vi$}}
    \psfrag{p}[][]{\scriptsize\textsf{$p$}}
    \psfrag{v}[][]{\scriptsize\textsf{$v$}}
    \psfrag{K}[][]{\hspace*{-0.2cm}\scriptsize{\textsf{$K$}}}
    \psfrag{\%/100}[][]{\scriptsize{\textsf{$\%\cdot 100$}}}
    \psfrag{0}[][]{\tiny{\textsf{0}}}
    \psfrag{0.1}[][]{\tiny{\textsf{0.1}}}
    \psfrag{0.2}[][]{\tiny{\textsf{0.2}}}
    \psfrag{0.3}[][]{\tiny{\textsf{0.3}}}
    \psfrag{0.4}[][]{\tiny{\textsf{0.4}}}
    \psfrag{0.5}[][]{\tiny{\textsf{0.5}}}
    \psfrag{0.6}[][]{\tiny{\textsf{0.6}}}
    \psfrag{0.7}[][]{\tiny{\textsf{0.7}}}
    \psfrag{0.8}[][]{\tiny{\textsf{0.8}}}
    \psfrag{0.9}[][]{\tiny{\textsf{0.9}}}
    \psfrag{1}[][]{\tiny{\textsf{1}}}
    \psfrag{-3}[][]{\tiny{\textsf{-3}}}
    \psfrag{-4}[][]{\tiny{\textsf{-4}}}

    \psfrag{0.01}[][]{\tiny\textsf{0.01}}
    \psfrag{0.02}[][]{\tiny\textsf{0.02}}
    \psfrag{0.04}[][]{\tiny\textsf{0.04}}
    \psfrag{0.06}[][]{\tiny\textsf{0.06}}
    \psfrag{0.78}[][]{\tiny\textsf{0.78}}
    \psfrag{0.82}[][]{\tiny\textsf{0.82}}
    \psfrag{0.84}[][]{\tiny\textsf{0.84}}
    \psfrag{5}[][]{\tiny{\textsf{5}}}
    \psfrag{10}[][]{\tiny{\textsf{10}}}
    \psfrag{15}[][]{\tiny{\textsf{15}}}

        \includegraphics[width=0.8\textwidth]{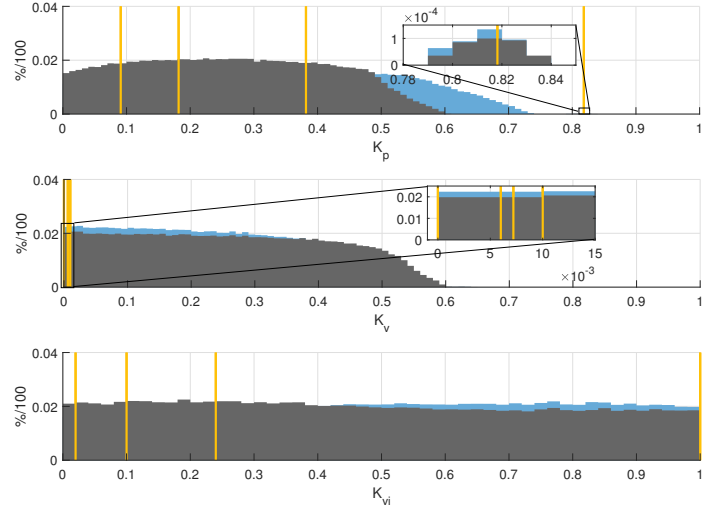}
        \caption
        {(Color online) The final safe set obtained from SafeOpt if Algorithm \ref{alg:InitiChoice} was used with $m_0=100$ (blue) and if a fixed starting point for the expander search was used (black). The initial safe set $S_0$ is marked with yellow}
        \label{fig:SafeSetsComparison100100}
\end{figure}

\begin{figure}[!tbp]
    \centering
        \psfrag{vi}[][]{\scriptsize\textsf{$vi$}}
    \psfrag{p}[][]{\scriptsize\textsf{$p$}}
    \psfrag{v}[][]{\scriptsize\textsf{$v$}}
    \psfrag{K}[][]{\hspace*{-0.2cm}\scriptsize{\textsf{$K$}}}
    \psfrag{\%/100}[][]{\scriptsize{\textsf{$\%\cdot 100$}}}
    \psfrag{0}[][]{\tiny{\textsf{0}}}
    \psfrag{0.1}[][]{\tiny{\textsf{0.1}}}
    \psfrag{0.2}[][]{\tiny{\textsf{0.2}}}
    \psfrag{0.3}[][]{\tiny{\textsf{0.3}}}
    \psfrag{0.4}[][]{\tiny{\textsf{0.4}}}
    \psfrag{0.5}[][]{\tiny{\textsf{0.5}}}
    \psfrag{0.6}[][]{\tiny{\textsf{0.6}}}
    \psfrag{0.7}[][]{\tiny{\textsf{0.7}}}
    \psfrag{0.8}[][]{\tiny{\textsf{0.8}}}
    \psfrag{0.9}[][]{\tiny{\textsf{0.9}}}
    \psfrag{1}[][]{\tiny{\textsf{1}}}
    \psfrag{-3}[][]{\tiny{\textsf{-3}}}
    \psfrag{-4}[][]{\tiny{\textsf{-4}}}

    \psfrag{1}[][]{\tiny{\textsf{1}}}
    \psfrag{2}[][]{\tiny{\textsf{2}}}

    \psfrag{0.01}[][]{\tiny\textsf{0.01}}
    \psfrag{0.02}[][]{\tiny\textsf{0.02}}
    \psfrag{0.04}[][]{\tiny\textsf{0.04}}
    \psfrag{0.06}[][]{\tiny\textsf{0.06}}
    \psfrag{0.78}[][]{\tiny\textsf{0.78}}
    \psfrag{0.82}[][]{\tiny\textsf{0.82}}
    \psfrag{0.84}[][]{\tiny\textsf{0.84}}
    \psfrag{5}[][]{\tiny{\textsf{5}}}
    \psfrag{10}[][]{\tiny{\textsf{10}}}
    \psfrag{15}[][]{\tiny{\textsf{15}}}

        \includegraphics[width=0.8\textwidth]{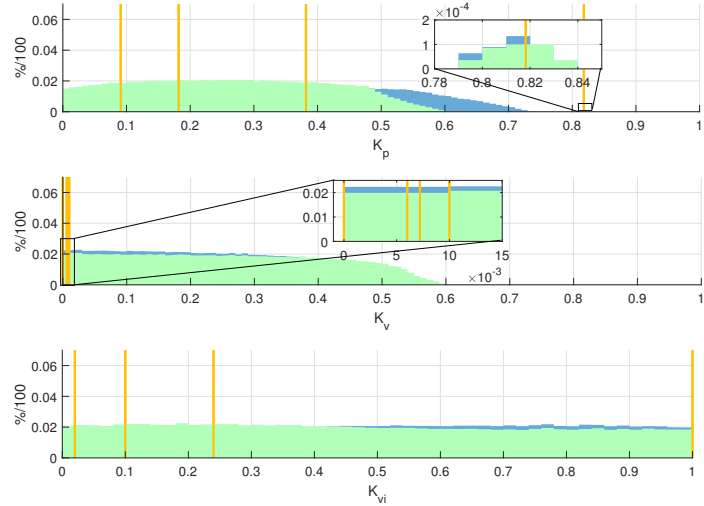}
        \caption{(Color online) The final safe set obtained from SafeOpt if Algorithm \ref{alg:InitiChoice} was used with $m_0=100$  generated with Latin hypercube sampling (blue) and with random sampling (light green). The initial safe set $S_0$ is marked with yellow}
        \label{fig:SafeSetsComparison100100rand}
\end{figure}

Figures \ref{fig:SafeSetsComparison100100} and \ref{fig:SafeSetsComparison100100rand} show a comparison of the safe set obtained for 100 points if the Latin hypercube sampling was used with a fixed initial guess (black in Fig. \ref{fig:SafeSetsComparison100100}) and with a random sampling (light green in Fig. \ref{fig:SafeSetsComparison100100rand}). In all the cases, the safe sets cover a similar part of the search space, around 30\% (first and last row in Table \ref{tbl:resultsAlgorithm}). The main difference is in the time necessary to find a solution (last column in Table \ref{tbl:resultsAlgorithm}). The algorithm proposed in the paper allows obtaining the same safe set while being faster. Using Latin hypercube sampling allows better coverage of the search space, leading to a speed-up of 0.8 s compared to random sampling. Thus, the analysis of the safe sets indicates that there is potential in exploiting the initial point for the expander search by adjusting both $m_0$ and the sampling method, to provide risk assessment with respect to parameters of the chosen controller.

\section{Optimization of the control parameters of a high-precision motion system}
\label{sec:ResultsArgus}
\begin{figure}[!tbp]
\centering
\includegraphics[width=0.6\textwidth]{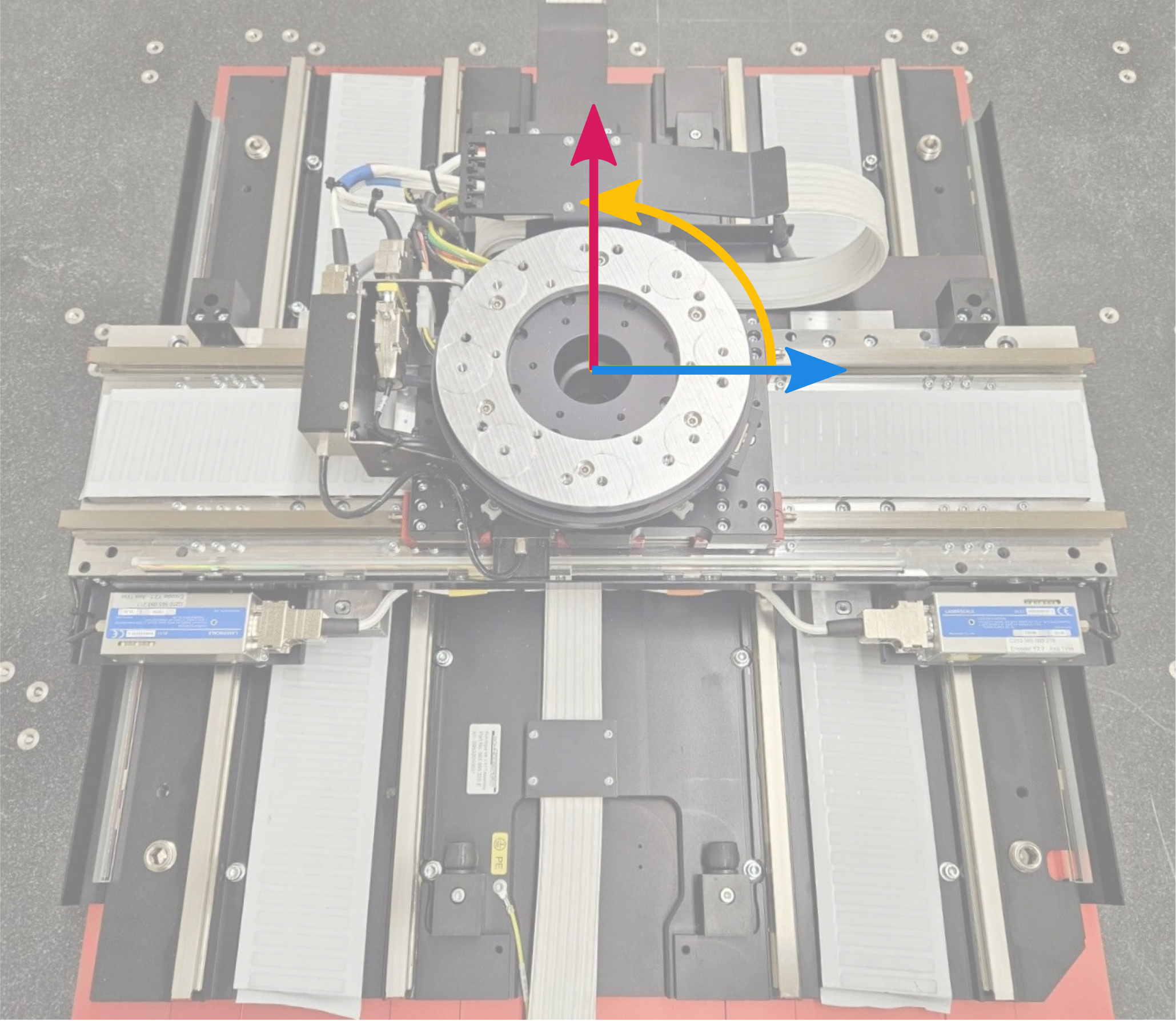}
\caption[Precision motion system]{High precision motion system with the two linear axes (lower (pink), and upper (blue) that is considered in this work), and the rotational axis (yellow)}
\label{fig:argus_system}
\end{figure}

\begin{figure}[!tbp]
    \centering
\scalebox{0.9}{
\scriptsize
    \begin{tikzpicture}[auto, node distance=2cm]
    \node [input, name=input] {};
    \node [sum, right of=input] (sum) {};
    \node [left of=sum, node distance=1cm](tmpIn){};
    \draw (tmpIn) to[short, -*] (tmpIn);
    \node [block, right of=sum,node distance=2cm, text width=2cm, align=center] (p_controller) {\textsf{\scriptsize{Position controller}}};
    \node [block, right of=p_controller,
            node distance=5cm, text width=2cm, align=center] (v_controller) {\scriptsize{\textsf{Velocity controller}}};
    \node[block, right of=v_controller, node distance=4cm](system){\scriptsize\textsf{System}};
    \node[block, above of=p_controller, text width=2.1cm, align=center](feedforward_v)
    {\textsf{\scriptsize{Feedforward velocity}}};
    \node [right of=feedforward_v, node distance=2cm](tmpIFF){};
    \draw (tmpIFF) to[short, -*] (tmpIFF);
    \node[block, above of=v_controller, text width=2.2cm, align=center](feedforward_a){\textsf{\scriptsize{Feedforward acceleration}}};
    \node[sum, right of=p_controller, node distance=2cm](sum1){};
    \node[sum, left of=v_controller, node distance=2cm](sum2){};
    \node[sum, left of=system, node distance=1.8cm](sum3){};
    \draw [->,>=Stealth] (sum) -- node {\scriptsize$p_e$} (p_controller);
    \draw[->,>=Stealth] (p_controller) -- node[pos=0.9] {} (sum1);
    \draw[->,>=Stealth] (sum1) --node {\scriptsize$v_r$} (sum2);
    \draw[->,>=Stealth] (sum2) --node {\scriptsize$v_e$} (v_controller);
    \draw[->,>=Stealth] (v_controller) --node[pos=0.95] {} (sum3);
    \draw[->,>=Stealth] (sum3) --node {\scriptsize$I_r$} (system);
    \draw[->,>=Stealth] (feedforward_v) -| node[pos=0.6] {\scriptsize VFF} (sum1);
    \draw[-] (feedforward_v) -| node[pos=0.95]{} (sum1);
    \draw[->,>=Stealth] (feedforward_v) -- node {} (feedforward_a);
    \draw[->,>=Stealth] (feedforward_a) -| node {\scriptsize AFF} (sum3);
    \draw[-] (feedforward_a) -| node [pos=0.95]{} (sum3);
    \draw[->,>=Stealth] (input) -- node[pos=0.95] {} (sum);
    \draw[-] (input) -- node{\scriptsize$p_r$} (sum);
    \draw[->,>=Stealth] (input) |- node {} (feedforward_v);
    \node[output, right of=system](output_p){};
    \draw[->,>=Stealth] (system) -- node {\scriptsize$p$}(output_p);
    \node[tmp, below of=system](tmp1){};
    \node[tmp, right of=system, node distance=1.5cm](tmp_output){};
    \draw (tmp_output) to[short, -*] (tmp_output);
    \draw[-] (tmp_output) |-node{}(tmp1);
    \draw[->,>=Stealth] (tmp1) -| node[pos=0.95]{$-$} (sum);
    \node[tmp, left of=p_controller, node distance=4cm] (tmp_input){};
    \draw[-] (sum) -- node {} (tmp_input);
    \node[output, below of=system, node distance=1cm](output_v){};
    \draw[-] (system) -- node{}(output_v);
    \draw[->,>=Stealth] (output_v) -| node[pos=0.95] {$-$}(sum2);
    \draw (output_v) to[short, -*] (output_v);
    \node[output, right of=output_v](tmp_v){};
    \draw[->,>=Stealth](output_v) -- node{\scriptsize$v$} (tmp_v);
    \end{tikzpicture}
    }
    \caption{Simplified block diagram of the controller. The signals $p_r, v_r, I_r$ are the reference signals of position, velocity and current respectively, $p_e$, $v_e$ are the position and velocity error signals from which the objective and constraint features are calculated, while $p$ and $v$ are the actual position and velocity. The signals VFF and AFF are feedforward signals to the velocity and acceleration cycle of the cascaded controller. The block ``Position controller'' is a proportional controller with parameter PKP, the ``Velocity controller'' is a proportional-integral (PI) controller with parameters VKP and VKI, the velocity feedforward gain is fixed, VFF=1, while the acceleration feedforward gain AFF is tuned}
    \label{fig:argus controller}    
\end{figure}
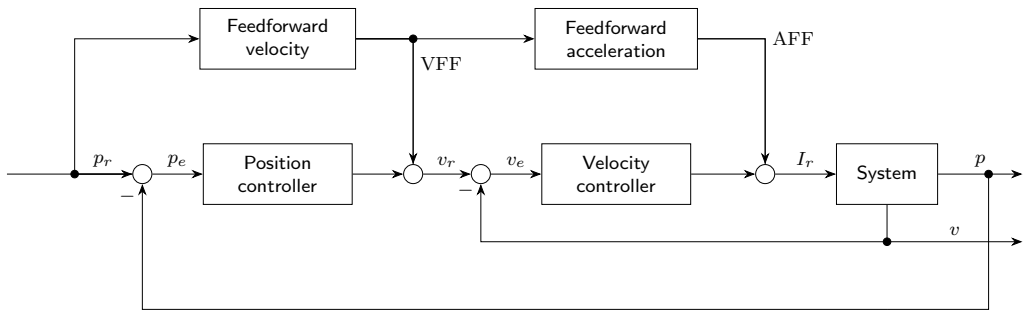

The experimental validation of the proposed grid-free SafeOpt was done by tuning the controller gains of a high-precision motion system (motion stage) from Schneeberger Linear Technology~\citep{rothfuss2022metalearning}. The system is shown in Fig. \ref{fig:argus_system} and consists of a 2D positioning stage with two orthogonal linear axes (pink and blue) and one rotational axis (yellow), though here we only consider the motion along the upper linear axis (blue). The axis is driven by a permanent magnet AC motor with precision encoders for position and speed tracking. The positioning accuracy of the axis is below 10 $\mu$m, with repeatability below 0.7 $\mu$m, and 3$\sigma$ stability below 1 nm. The system is controlled by a cascade controller with a proportional controller (P) for the position and a proportional-integral (PI) controller for the velocity (Fig. \ref{fig:argus controller}) that should be tuned to achieve sub-micrometer precision~\citep{lee2000intelligent}. Following~\cite{Safe_Koenig2023} we use a sampling time of the controller and the data acquisition of the system of 2.5 kHz. The buffer length of each measurement is 3000 points, which results in a 1.2 s measurement for each step. The performance of the system is given by the filtered average position error over the 1.2 s measurement. 

The results of SafeOpt in grid-based and grid-free version were compared to the gains of the automatic tuner that is built into the controller and the benchmark algorithm (Goal-oriented Safe Exploration) GoOSE developed by~\cite{könig2021safe}, adapted for continuous, adaptive controller tuning in precision motion systems by~\cite{Safe_Koenig2023}. GoOSE ensures that every input to the system satisfies an unknown, observable constraint. For controller tuning, it unifies time-varying Gaussian process bandit optimization from~\cite{bogunovic2016time} with multi-task Gaussian processes from~\cite{swersky2013multi}, and with efficient safe set search based on particle swarm optimization as introduced in~\cite{Safe_Koenig2023}.

\subsection{Controller tuning problem}
The gains of the controller are the proportional position gain (PKP), the proportional velocity gain (VKP), the integral velocity gain (VKI), and the feedforward gain of the acceleration (AFF). The step size $\alpha$, which is the distance the system covers in one motion, was a task parameter as described by~\cite{könig2021safe}. The goal was to minimize the average position error for one linear axis, and a constraint was put on the fast Fourier transform of the velocity error $e_v$ and the average position error. 
The optimization problem is formulated as:
\begin{subequations}
    \begin{align}
        \min_{x \in A}\quad  & e_{\text{avg}}(x), \label{eq:CostReal} \\
        \text{subject to:}\quad & e_{\text{avg}}(x) \leq J_{\max}^1(\alpha),\label{eq:eavg} \\
        & \max_{f \in [140\text{Hz}, 1250\text{Hz}]} |\text{FFT}\left[\xi(i, n_{s})\mathrm{v_e}(t_{i})\right](x)| \leq J_{\max}^2(\alpha)\label{eq:FFTmax}
    \end{align}
\end{subequations}
where:
\begin{equation}
    e_{\text{avg}}(x):=\frac{1}{n_{P} - n_{s}} {\sum\limits_{i=n_{s}}^{n_{P}} |\xi(i, n_{s})\mathrm{p_e}(t_{i})|}.
\end{equation}
The position and velocity error $\mathrm{p_e}$ and $\mathrm{v_e}$ depend on the controller parameters, $x$ is a vector with the four gains and the stepsize [PKP, VKP, VKI, AFF, $\alpha$] and $J_{\max}^1$ and $J_{\max}^2$ are the constraint limits which depend on the stepsize $\alpha$. The search space is defined as $A = [100,450]\times[450,1500]\times[800,2500]\times[0,2]$, and $\xi(i, n_{s})$ is the right-sided sigmoid filter function 
\[\xi(i, n_{s}) = 1 - (1+\mathrm{exp}(-(i-n_{s}-150)/10))^{-1}.
\]
Finally, $n_{s}$ is the time sample where the movement of the position reference function ends and the sampling time begins, while $n_{P}$ is the time sample of the end of the settling time set at 1.2 s after start of the movement.

\subsection{Results}
All learning algorithms, grid SafeOpt, grid-free SafeOpt, and GoOSE, used Gaussian processes to model~\eqref{eq:eavg} and~\eqref{eq:FFTmax} with a squared exponential kernel with the lengthscales: l\textsubscript{PKP} = 50, l\textsubscript{VKP} = 100, l\textsubscript{VKI} = 200, l\textsubscript{AFF} and l\textsubscript{stepsize} = 0.3. The variance of the kernel for $e_{\text{avg}}$ was set to $0.36\times 10^{-10}$ and for FFT\textsubscript{max} to $1\times 10^{-4}$.  We set $\beta=3$ in~\eqref{eq:Bounds}, as in practice choosing $\beta\geq 2$ often proves sufficient~\citep{könig2021safe}. The number of points in Algorithm \ref{alg:InitiChoice} was set $m=300$, and the samples were obtained using Latin hypercube sampling. The stopping criterion for all the algorithms was set to 100 iterations. As the local solver algorithm we chose a custom implementation of Generalised Pattern Search from~\cite{Derivative_Audet2017} in Python 3.9. The tests were run on a Dell Inc. Precision 5820 Tower PC with 64~GB~RAM.

\begin{figure}[!tbp]
\psfrag{Time [s]}[][]{\scriptsize\textsf{Time [s]}}
\psfrag{Iteration number}[][]{\scriptsize\textsf{Iteration number}}
\psfrag{GoOSe benchmark}[][]{\hspace{-0.09cm}\scriptsize\textsf{GoOSE}}
\psfrag{Grid-free SafeOpt}[][]{\scriptsize\textsf{Grid-free SafeOpt}}
\psfrag{Grid SafeOpt}[][]{\hspace{0.09cm}\scriptsize\textsf{Grid SafeOpt}}
\psfrag{10}[][]{\tiny\textsf{10}}
\psfrag{25}[][]{\tiny\textsf{25}}
\psfrag{50}[][]{\tiny\textsf{50}}
\psfrag{75}[][]{\tiny\textsf{75}}
\psfrag{100}[][]{\tiny\textsf{100}}
\psfrag{0}[][]{\tiny\textsf{0}}
\psfrag{1}[][]{\tiny\textsf{1}}
\psfrag{2}[][]{\tiny\textsf{2}}
\psfrag{3}[][]{\tiny\textsf{3}}

\psfrag{Grid-free SO optimizers}[][]{\scriptsize\textsf{Grid-free SO optimizers}}
\psfrag{Grid-free SO expanders}[][]{\hspace{-0.09cm}\scriptsize\textsf{Grid-free SO expanders}}
\psfrag{Grid SO optimizers}[][]{\scriptsize\textsf{Grid SO optimizers}}
\psfrag{Grid SO expanders}[][]{\hspace{-0.09cm}\scriptsize\textsf{Grid SO expanders}}

    \centering
    \begin{subfigure}[t]{0.75\textwidth}
        \includegraphics[width=\textwidth]{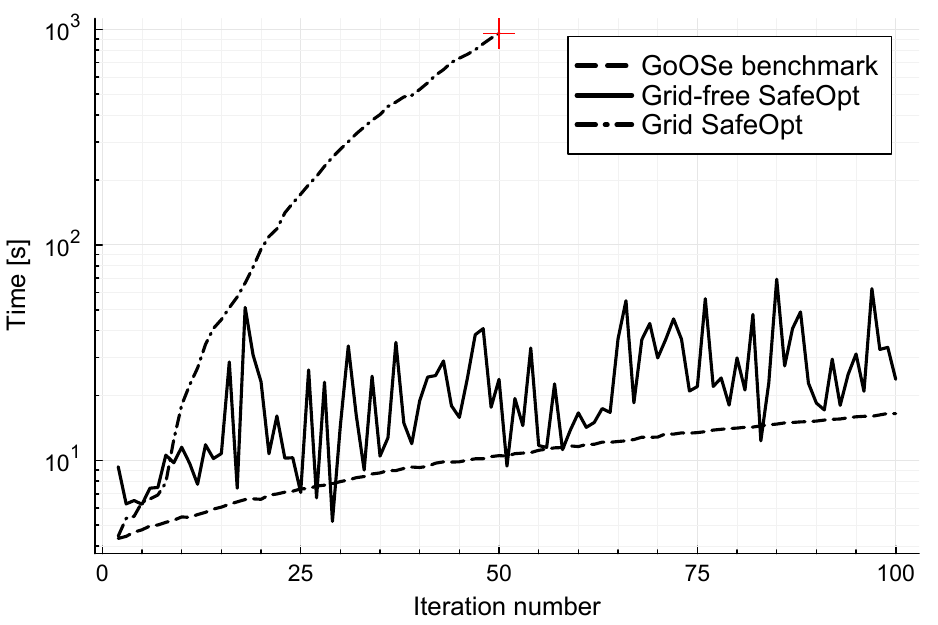}
        \caption[Iteration time comparison]
        {The iteration times for grid SafeOpt (dash-dotted) and grid-free SafeOpt (solid), compared to the benchmark GoOSE (dashed)}
        \label{fig:comparison_iteration_time}
    \end{subfigure}%
    \hfill
    \begin{subfigure}[t]{0.75\textwidth}
        \includegraphics[width=\textwidth]{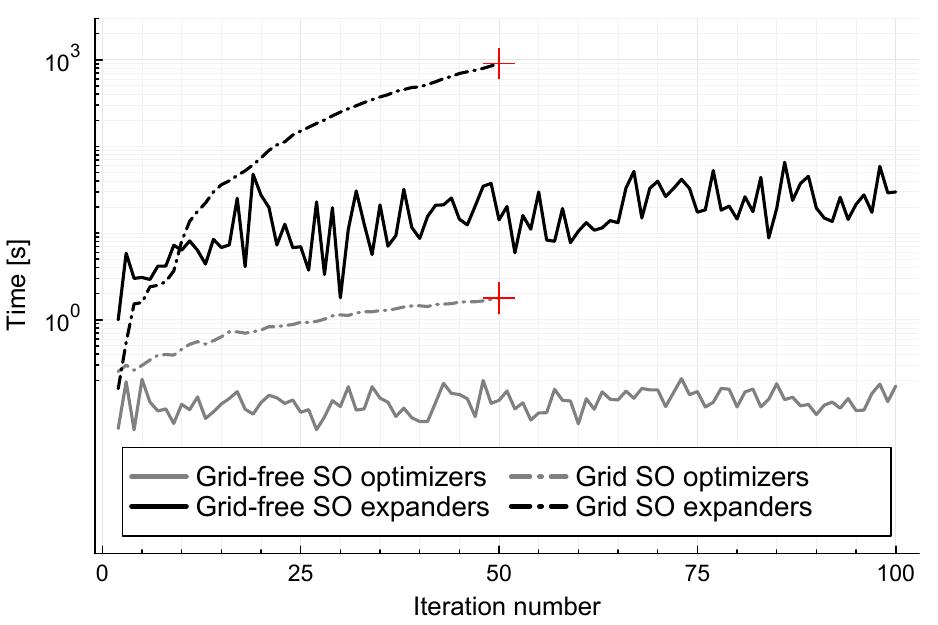}
        \caption{The iteration times for expanders (black) and optimizers (gray) from grid SafeOpt (dash-dotted) and grid-free SafeOpt (solid)}
        \label{fig:comparison_search_time}
    \end{subfigure}%
    \caption{Time comparison}
    \label{fig:time_comparison}
\end{figure}

The runtime of a single iteration is shown in Fig. \ref{fig:comparison_iteration_time}. While the iteration times for both SafeOpt algorithms start around 10 s, the iteration time of grid SafeOpt grows much faster than grid-free SafeOpt. As a result, grid SafeOpt was interrupted after 4 hours (50 iterations) since iteration times reached 20 minutes per iteration. Conversely, an iteration of grid-free SafeOpt performs similarly to the benchmark GoOSE. Thus, grid-free SafeOpt and GoOSE were run for 100 iterations taking 37 and 18 minutes, respectively. The difference in timings between GoOSE and grid-free SafeOpt arises because GoOSE focuses on evaluating the optimizers, limiting the use of expanders and thus removing the computation of the auxiliary GPs from~\eqref{berkenkamp_next_point}.

Figure \ref{fig:comparison_search_time} shows the iteration times broken down into the optimizer search time and the expander search time. For both SafeOpt algorithms, most of the iteration time is taken up by the expander search. The time for the optimizer search is similar for both. The long search times for expanders stems from the calculation of $u_{n,j,(x,l_n(x,j))}(x')$ since this is done by adding and removing $(x,l_n(x,j))$ from the GPs. The computational cost grows with the number of evaluations added to the auxiliary GPs (line 7 in Algorithm \ref{alg:SafeOpt}). Therefore, the cost also grows with the number of iterations, since an evaluation is added in every iteration. In grid SafeOpt these calculations have to be done for every safe point in the entire grid, leading to a long runtime. In grid-free SafeOpt the calculation of necessary to find $\mathcal{E}_j$ in the expander search is only done for the points pattern search chooses to evaluate. As a result, the time for a single iteration in grid-free SafeOpt remains similar over time.

\begin{figure*}[!tbp]
    \centering
\psfrag{Iteration number}[][]{\scriptsize\textsf{Iteration number}}

    \psfrag{GoOSe benchmark}[][]{\hspace{0.45cm}\scriptsize\textsf{GoOSE}}
\psfrag{Grid-free SafeOpt}[][]{\hspace{0.42cm}\scriptsize\textsf{Grid-free SafeOpt}}
\psfrag{Grid SafeOpt}[][]{\hspace{0.25cm}\scriptsize\textsf{Grid SafeOpt}}
\psfrag{10}[][]{\tiny\textsf{10}}
\psfrag{25}[][]{\tiny\textsf{25}}
\psfrag{50}[][]{\tiny\textsf{50}}
\psfrag{75}[][]{\tiny\textsf{75}}
\psfrag{100}[][]{\tiny\textsf{100}}
\psfrag{0}[][]{\tiny\textsf{0}}

\psfrag{175}[][]{\tiny\textsf{175}}
\psfrag{200}[][]{\tiny\textsf{200}}
\psfrag{225}[][]{\tiny\textsf{225}}
\psfrag{250}[][]{\tiny\textsf{250}}
\psfrag{275}[][]{\tiny\textsf{275}}
\psfrag{300}[][]{\tiny\textsf{300}}
\psfrag{325}[][]{\tiny\textsf{325}}

\psfrag{600}[][]{\tiny\textsf{600}}
\psfrag{700}[][]{\tiny\textsf{700}}
\psfrag{800}[][]{\tiny\textsf{800}}
\psfrag{650}[][]{\tiny\textsf{650}}
\psfrag{750}[][]{\tiny\textsf{750}}
\psfrag{850}[][]{\tiny\textsf{850}}

\psfrag{900}[][]{\tiny\textsf{900}}
\psfrag{1000}[][]{\tiny\textsf{1000}}
\psfrag{1100}[][]{\tiny\textsf{1100}}
\psfrag{1200}[][]{\tiny\textsf{1200}}
\psfrag{1300}[][]{\tiny\textsf{1300}}
\psfrag{1400}[][]{\tiny\textsf{1400}}

\psfrag{0.00}[][]{\tiny\textsf{0.00}}
\psfrag{0.25}[][]{\tiny\textsf{0.25}}
\psfrag{0.50}[][]{\tiny\textsf{0.50}}
\psfrag{0.75}[][]{\tiny\textsf{0.75}}
\psfrag{1.00}[][]{\tiny\textsf{1.00}}

\psfrag{PKP}[][]{\scriptsize\textsf{PKP}}
\psfrag{VKI}[][]{\scriptsize\textsf{VKI}}
\psfrag{VKP}[][]{\scriptsize\textsf{VKP}}
\psfrag{AFF}[][]{\scriptsize\textsf{AFF}}

    \begin{subfigure}{0.49\textwidth}
        \includegraphics[width=\textwidth]{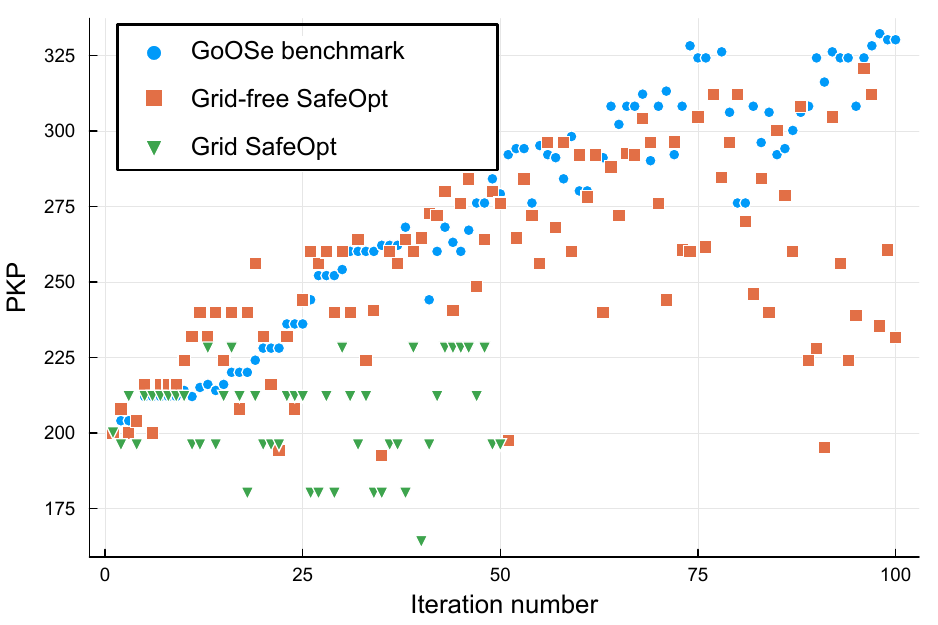}
        \caption{PKP}
    \end{subfigure}
    \begin{subfigure}{0.49\textwidth}
        \includegraphics[width=\textwidth]{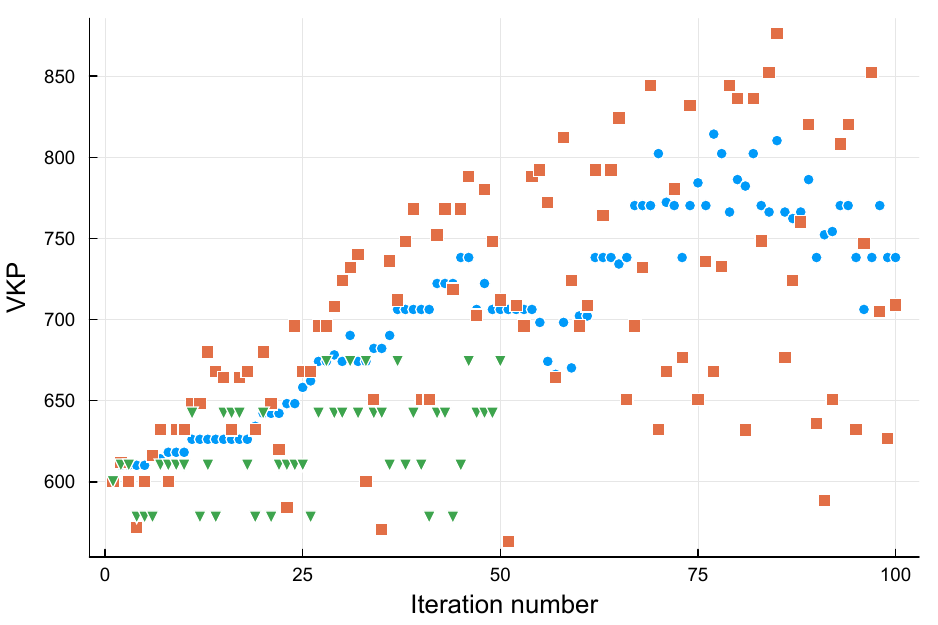}
        \caption{VKP}
    \end{subfigure}
    \hfill
    \begin{subfigure}{0.49\textwidth}
        \includegraphics[width=\textwidth]{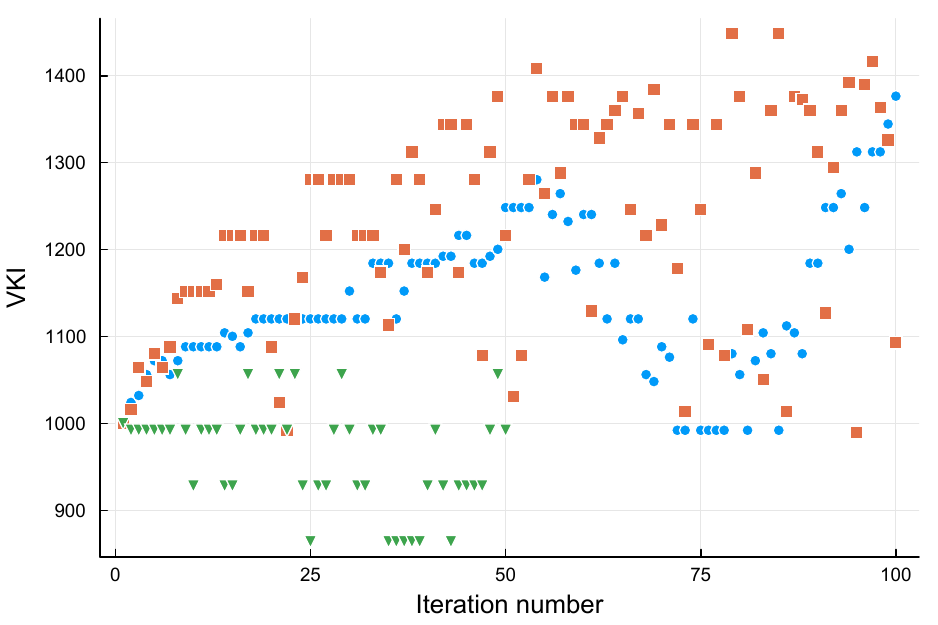}
        \caption{VKI}
    \end{subfigure}
    \begin{subfigure}{0.49\textwidth}
        \includegraphics[width=\textwidth]{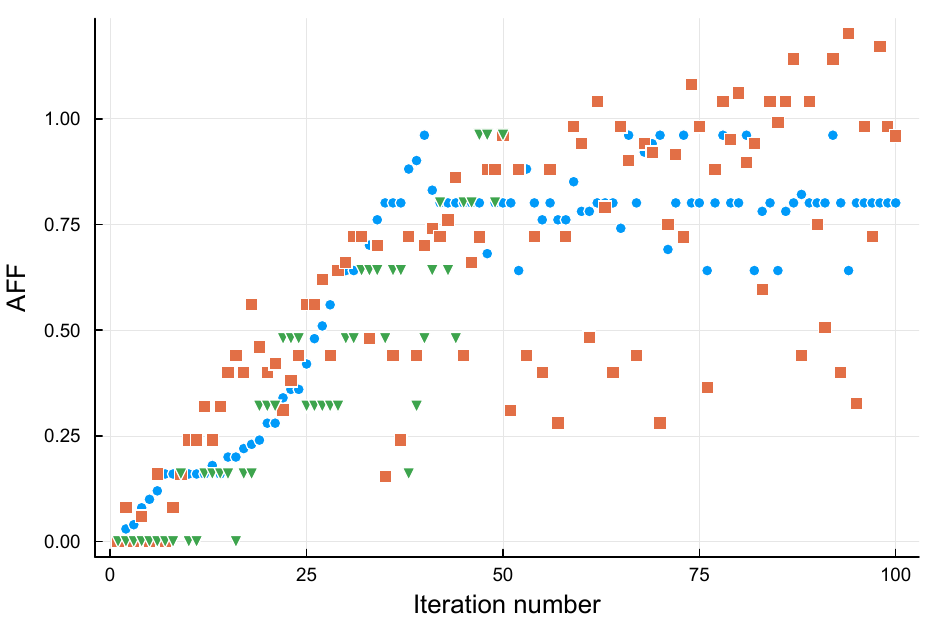}
        \caption{AFF}
    \end{subfigure}
    \caption{Comparison of the evaluated points from grid SafeOpt (triangles) and grid-free SafeOpt (squares) with the benchmark solution from GoOSE (circles)}
    \label{fig:eval_points_comparison}
\end{figure*}

\begin{figure}[tbp]
    \centering
    \psfrag{Iteration number}[][]{\scriptsize\textsf{Iteration number}}
    \psfrag{Error [mm]}[][]{\scriptsize\textsf{Error [{\textmu}m]}}

    \psfrag{GoOSe benchmark}[][]{\hspace{-0.09cm}\scriptsize\textsf{GoOSE}}
\psfrag{Grid-free SafeOpt}[][]{\hspace{0.0cm}\scriptsize\textsf{Grid-free SafeOpt}}
\psfrag{Grid SafeOpt}[][]{\hspace{0.09cm}\scriptsize\textsf{Grid SafeOpt}}
\psfrag{10}[][]{\tiny\textsf{10}}
\psfrag{25}[][]{\tiny\textsf{25}}
\psfrag{50}[][]{\tiny\textsf{50}}
\psfrag{75}[][]{\tiny\textsf{75}}
\psfrag{100}[][]{\tiny\textsf{100}}
\psfrag{0}[][]{\tiny\textsf{0}}

\psfrag{1.0\32710}[][]{\tiny\textsf{1.0$\times$10}}
\psfrag{2.0\32710}[][]{\tiny\textsf{2.0$\times$10}}
\psfrag{3.0\32710}[][]{\tiny\textsf{3.0$\times$10}}
\psfrag{4.0\32710}[][]{\tiny\textsf{4.0$\times$10}}
\psfrag{5}[][]{\tiny\textsf{-2}}

    \includegraphics[width=0.75\textwidth]{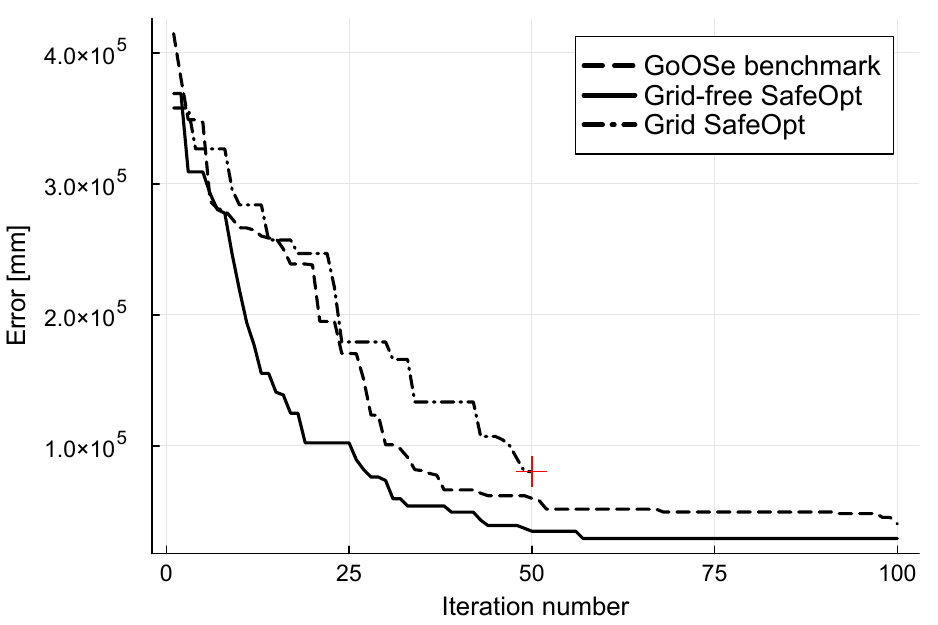}
    \caption[Comparison of the smallest $e_{avg}$ evaluated]
    {Comparison of the smallest $e_{\text{avg}}$ as a function of iteration for grid SafeOppt (dash-dotted), grid-free SafeOpt (solid), and the benchmark GoOSE (dashed)}
    \label{fig:smallest_err_comparison}
\end{figure}

The points evaluated by the three algorithms are shown in Fig. \ref{fig:eval_points_comparison}. Both SafeOpt algorithms are searching in the same part of the search space, but grid SafeOpt (triangles) evaluates points that are on a grid while the points selected by grid-free SafeOpt (squares) are less restricted. 
Grid-free SafeOpt finds points with a smaller $e_{\text{avg}}$ compared to grid SafeOpt (Fig. \ref{fig:smallest_err_comparison}). This is because grid SafeOpt is restricted to the grid and thus unable to find a better point that lies between points of the grid.
Finally, as GoOSE computes only optimizers, the values of the parameters are less spread across the search space. Grid-free SafeOpt explores a wider part of the search space, at the expense of increased computational time.

\subsection{Comparison with automatic tuner and GoOSE}

\begin{table*}[]
\centering
\scriptsize\sffamily
\caption{Comparison of grid-free SafeOpt with grid SafeOpt, automatic tuning, and benchmark GoOSE using a 10 mm step on 10 different positions on the motion stage, and the mean and standard deviation of the cost over the 10 repetitions over the positions are included. The superscript \textcolor{red}{$^{+}$} indicates that the solution was interrupted after 50 iterations. The cost is uniformly scaled by $10^{-7}$ for presentation}
\label{tab:prediction_comparison}
\resizebox{\textwidth}{!}{%
\begin{tabular}{@{}lccc@{}}
\toprule
Algorithm         & Solution $x^*$ = {[}PKP, VKP, VKI, AFF{]} & Scaled Cost~\eqref{eq:CostReal}          & Runtime {[}min{]}                         \\ \midrule
Grid SafeOpt      & {[}212.1, 546.1, 800, 0.961{]}            & 136.7 $\pm$ 8.81   & 253\textcolor{red}{$^{+}$}  \\
Grid-free SafeOpt & {[}304.1, 836.2, 1440.4, 0.841{]}         & 23.29 $\pm$ 3.44   & 37                                        \\
Autotuning               & {[}350, 600, 2000, 0{]}                   & 52.63 $\pm$ 4.15 & 2                                         \\
GoOSE  & {[}332.2, 746.3, 1344.5, 0.741{]} & 31.65 $\pm$ 3.98 & 18                                         \\\bottomrule
\end{tabular}%
}
\end{table*}

Table \ref{tab:prediction_comparison} shows the predicted optimal gains and the resulting average position error for grid SafeOpt, grid-free SafeOpt, and GoOSE. Grid-free SafeOpt predicted a better optimum than grid SafeOpt within the given timeframe. Figure \ref{fig:smallest_err_comparison} shows that grid-free Safeopt evaluates points with a smaller average position error than grid SafeOpt. The proposed grid-free algorithm also achieves performance 30\% better than GoOSE, thanks to using expanders as well as optimizers and exploring a larger part of the search space.

\begin{figure}[!tbp]
    \centering
        \psfrag{Time [ms]}[][]{\scriptsize\textsf{Time [ms]}}
    \psfrag{Average position error [mm]}[][]{\scriptsize\textsf{Average position error [{\textmu}m]}}

    \psfrag{GoOSe benchmark}[][]{\hspace{-0.09cm}\scriptsize\textsf{GoOSE}}
\psfrag{Grid-free SafeOpt}[][]{\hspace{0.0cm}\scriptsize\textsf{Grid-free SafeOpt}}
\psfrag{Grid SafeOpt}[][]{\hspace{0.09cm}\scriptsize\textsf{Grid SafeOpt}}
\psfrag{Autotuning}[][]{\hspace{0.1cm}\scriptsize\textsf{Autotuning}}

\psfrag{50}[][]{\tiny\textsf{50}}
\psfrag{100}[][]{\tiny\textsf{100}}
\psfrag{0}[][]{\tiny\textsf{0}}
\psfrag{150}[][]{\tiny\textsf{150}}
\psfrag{200}[][]{\tiny\textsf{200}}
\psfrag{250}[][]{\tiny\textsf{250}}

\psfrag{0.002}[][]{\tiny\textsf{2.0}}
\psfrag{0.001}[][]{\tiny\textsf{1.0}}
\psfrag{0.000}[][]{\tiny\textsf{0.0}}
\psfrag{-0.001}[][]{\hspace{0.09cm}\tiny\textsf{-1.0}}

    \includegraphics[width=0.75\textwidth]{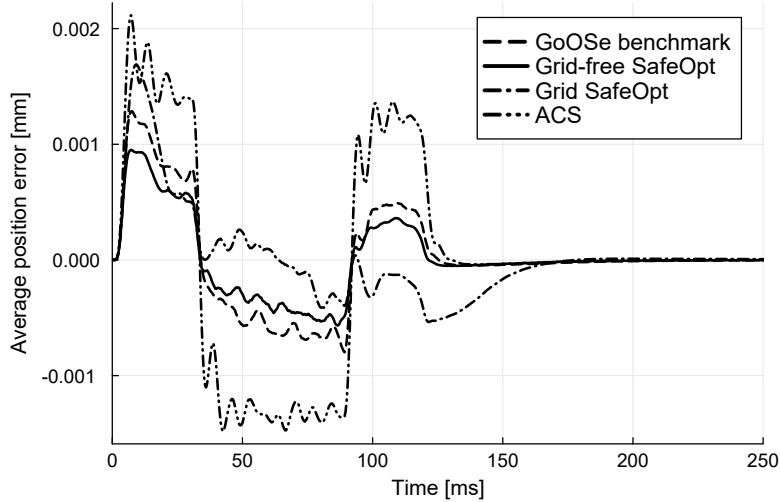}
    \caption{Position error comparison averaged over 10 runs, obtained for the optimal $x^*$ from Table \ref{tab:prediction_comparison}}
    \label{fig:argus_pos_err}
\end{figure}

Figure \ref{fig:argus_pos_err} shows the average position error for the four different configurations of the controller from Table \ref{tab:prediction_comparison}. The four controllers were validated for 10 different set-points and the average error is shown in Fig. \ref{fig:argus_pos_err}. All four algorithms drive the error towards zero. In particular, the proposed grid-free SafeOpt (solid) has better performance with the benchmark solution from GoOSE (dashed), thanks to using expanders to explore the search space. The amplitude with the configuration of the autotuning (double dash-dotted) is the biggest, while the gains found by grid-free SafeOpt result in the smallest amplitude. The solution of grid SafeOpt (single dash-dotted) resulted in a smaller amplitude than the autotuning but bigger than the configuration found by grid-free SafeOpt.

We also note that the grid-free SafeOpt reaches the optimum error in iteration 60. This indicates the potential of using a stopping criterion based on convergence instead of number of iterations~\citep{zagorowska2022efficient}. We also note that grid-free SafeOpt reached zero error at the same time as GoOSE, around 120 ms, whereas the autotuning needed 160 ms. This result confirms the advantages of using an optimization-based tuning if high precision is required.

\section{Conclusions and future work}
\label{sect:concl}

Learning-based controller tuning allows adjusting the parameters that satisfy chosen performance criteria while satisfying safety constraints. In this work, we present a new approach to safe learning, formulating the SafeOpt algorithm as a series of local optimization problems, thus avoiding exhaustive search and improving its computational performance. We also develop a method for initializing the local optimization problems to ensure their feasibility, while preserving the properties of SafeOpt, thus enabling controller tuning without explicitly formulating the optimization problem as a function of the controller parameters.

The proposed grid-free SafeOpt algorithm has been first validated in a simulation of cascade controller tuning, showing the impact of initialization on local solvers and confirming improved computational performance. In particular, the initialization can be used to adjust the safe sets obtained during the optimization, thus improving the flexibility of the grid-free SafeOpt. We then demonstrate experimentally the performance of the algorithm for controller tuning in a precision motion system. A comparison with the default autotuner shows the benefits of using optimization-based tuning to achieve the required sub-micrometer precision. The nearly seven-fold improvement in run-time compared to grid SafeOpt is achieved thanks to limiting the number of points during the search for the next iterate. 

The experiments also show that the efficient SafeOpt implementation achieved 30\% better tracking performance than a state-of-the-art benchmark algorithm, at the expense of a twice as long computational time. For this project, we used pattern search to solve the optimization problems in the grid-free SafeOpt. In the future, other optimization algorithms could potentially be used to further improve the run-time of the algorithm, which would enable its application in continuous optimization for use in time-varying systems.

\section*{Declaration of competing interest}
The authors declare that they have no known competing financial interests or personal relationships that could have appeared to influence the work reported in this paper.

\section*{Acknowledgement}

Research supported by NCCR Automation, a National Centre of Competence in Research, funded by the Swiss National Science Foundation (grant no. 180545), and by the European Research Council (ERC) under the H2020 Advanced Grant no. 787845 (OCAL). Marta Zagorowska also acknowledges funding from the Marie Curie Horizon Postdoctoral Fellowship project RELIC (grant no 101063948).

  \bibliographystyle{elsarticle-harv} 
\bibliography{biblio.bib}

\begin{thebibliography}{29}
\expandafter\ifx\csname natexlab\endcsname\relax\def\natexlab#1{#1}\fi
\providecommand{\url}[1]{\texttt{#1}}
\providecommand{\href}[2]{#2}
\providecommand{\path}[1]{#1}
\providecommand{\DOIprefix}{doi:}
\providecommand{\ArXivprefix}{arXiv:}
\providecommand{\URLprefix}{URL: }
\providecommand{\Pubmedprefix}{pmid:}
\providecommand{\doi}[1]{\href{http://dx.doi.org/#1}{\path{#1}}}
\providecommand{\Pubmed}[1]{\href{pmid:#1}{\path{#1}}}
\providecommand{\bibinfo}[2]{#2}
\ifx\xfnm\relax \def\xfnm[#1]{\unskip,\space#1}\fi
\bibitem[{{\AA}str{\"o}m and H{\"a}gglund(2006)}]{Advanced_Astroem2006}
\bibinfo{author}{{\AA}str{\"o}m, K.J.}, \bibinfo{author}{H{\"a}gglund, T.},
  \bibinfo{year}{2006}.
\newblock \bibinfo{title}{Advanced {PID} Control}.
\newblock \bibinfo{publisher}{ISA-The Instrumentation, Systems, and Automation
  Society}.
\bibitem[{Audet and Hare(2017)}]{Derivative_Audet2017}
\bibinfo{author}{Audet, C.}, \bibinfo{author}{Hare, W.}, \bibinfo{year}{2017}.
\newblock \bibinfo{title}{Derivative-free and blackbox optimization}.
\newblock Springer Series in Operations Research and Financial Engineering,
  \bibinfo{publisher}{Springer Cham}.
\bibitem[{Azizsoltani and Sadeghi(2018)}]{azizsoltani2018adaptive}
\bibinfo{author}{Azizsoltani, H.}, \bibinfo{author}{Sadeghi, E.},
  \bibinfo{year}{2018}.
\newblock \bibinfo{title}{Adaptive sequential strategy for risk estimation of
  engineering systems using {G}aussian process regression active learning}.
\newblock \bibinfo{journal}{Engineering Applications of Artificial
  Intelligence} \bibinfo{volume}{74}, \bibinfo{pages}{146--165}.
\bibitem[{Berkenkamp et~al.(2021)Berkenkamp, Krause and
  Schoellig}]{BayesianBerkenkamp2021}
\bibinfo{author}{Berkenkamp, F.}, \bibinfo{author}{Krause, A.},
  \bibinfo{author}{Schoellig, A.P.}, \bibinfo{year}{2021}.
\newblock \bibinfo{title}{Bayesian optimization with safety constraints: {S}afe
  and automatic parameter tuning in robotics}.
\newblock \bibinfo{journal}{Machine Learning} , \bibinfo{pages}{1--35}.
\bibitem[{Berkenkamp et~al.(2016)Berkenkamp, Schoellig and
  Krause}]{Berkenkamp_2016}
\bibinfo{author}{Berkenkamp, F.}, \bibinfo{author}{Schoellig, A.P.},
  \bibinfo{author}{Krause, A.}, \bibinfo{year}{2016}.
\newblock \bibinfo{title}{Safe controller optimization for quadrotors with
  {G}aussian processes}, in: \bibinfo{booktitle}{2016 IEEE International
  Conference on Robotics and Automation (ICRA)}, \bibinfo{publisher}{IEEE
  Press}. p. \bibinfo{pages}{491–496}.
\bibitem[{Bichon et~al.(2011)Bichon, McFarland and
  Mahadevan}]{bichon2011efficient}
\bibinfo{author}{Bichon, B.J.}, \bibinfo{author}{McFarland, J.M.},
  \bibinfo{author}{Mahadevan, S.}, \bibinfo{year}{2011}.
\newblock \bibinfo{title}{Efficient surrogate models for reliability analysis
  of systems with multiple failure modes}.
\newblock \bibinfo{journal}{Reliability {E}ngineering \& {S}ystem {S}afety}
  \bibinfo{volume}{96}, \bibinfo{pages}{1386--1395}.
\bibitem[{Bogunovic et~al.(2016)Bogunovic, Scarlett and
  Cevher}]{bogunovic2016time}
\bibinfo{author}{Bogunovic, I.}, \bibinfo{author}{Scarlett, J.},
  \bibinfo{author}{Cevher, V.}, \bibinfo{year}{2016}.
\newblock \bibinfo{title}{Time-varying {G}aussian process bandit optimization},
  in: \bibinfo{booktitle}{Artificial Intelligence and Statistics},
  \bibinfo{organization}{PMLR}. pp. \bibinfo{pages}{314--323}.
\bibitem[{Duivenvoorden et~al.(2017)Duivenvoorden, Berkenkamp, Carion, Krause
  and Schoellig}]{Constrained_Duivenvoorden2017}
\bibinfo{author}{Duivenvoorden, R.R.P.R.}, \bibinfo{author}{Berkenkamp, F.},
  \bibinfo{author}{Carion, N.}, \bibinfo{author}{Krause, A.},
  \bibinfo{author}{Schoellig, A.P.}, \bibinfo{year}{2017}.
\newblock \bibinfo{title}{Constrained {B}ayesian optimization with particle
  swarms for safe adaptive controller tuning}.
\newblock \bibinfo{journal}{{IFAC}-{PapersOnLine}} \bibinfo{volume}{50},
  \bibinfo{pages}{11800--11807}.
\bibitem[{Fauriat and Gayton(2014)}]{fauriat2014ak}
\bibinfo{author}{Fauriat, W.}, \bibinfo{author}{Gayton, N.},
  \bibinfo{year}{2014}.
\newblock \bibinfo{title}{{AK-SYS}: an adaptation of the {AK-MCS} method for
  system reliability}.
\newblock \bibinfo{journal}{Reliability Engineering \& System Safety}
  \bibinfo{volume}{123}, \bibinfo{pages}{137--144}.
\bibitem[{Fiducioso et~al.(2019)Fiducioso, Curi, Schumacher, Gwerder and
  Krause}]{Safe_Fiducioso2019}
\bibinfo{author}{Fiducioso, M.}, \bibinfo{author}{Curi, S.},
  \bibinfo{author}{Schumacher, B.}, \bibinfo{author}{Gwerder, M.},
  \bibinfo{author}{Krause, A.}, \bibinfo{year}{2019}.
\newblock \bibinfo{title}{Safe contextual {B}ayesian optimization for
  sustainable room temperature {PID} control tuning}, in:
  \bibinfo{booktitle}{Proceedings of the Twenty-Eighth International Joint
  Conference on Artificial Intelligence, {IJCAI-19}},
  \bibinfo{publisher}{International Joint Conferences on Artificial
  Intelligence Organization}. pp. \bibinfo{pages}{5850--5856}.
\bibitem[{Fujimoto et~al.(2022)Fujimoto, Sato and
  Nagahara}]{Controller_Fujimoto2022}
\bibinfo{author}{Fujimoto, Y.}, \bibinfo{author}{Sato, H.},
  \bibinfo{author}{Nagahara, M.}, \bibinfo{year}{2022}.
\newblock \bibinfo{title}{Controller tuning with {B}ayesian optimization and
  its acceleration: Concept and experimental validation}.
\newblock \bibinfo{journal}{Asian Journal of Control} \bibinfo{volume}{25},
  \bibinfo{pages}{2408--2414}.
\bibitem[{Iman and Shortencarier(1984)}]{osti_7091452}
\bibinfo{author}{Iman, R.L.}, \bibinfo{author}{Shortencarier, M.J.},
  \bibinfo{year}{1984}.
\newblock \bibinfo{title}{Fortran 77 program and user's guide for the
  generation of {L}atin hypercube and random samples for use with computer
  models} \URLprefix \url{https://www.osti.gov/biblio/7091452}.
  \bibinfo{note}{accessed: 26 Sep 2023}.
\bibitem[{Khosravi et~al.(2020)Khosravi, Behrunani, Smith, Rupenyan and
  Lygeros}]{CascadeKhosravi2020}
\bibinfo{author}{Khosravi, M.}, \bibinfo{author}{Behrunani, V.},
  \bibinfo{author}{Smith, R.S.}, \bibinfo{author}{Rupenyan, A.},
  \bibinfo{author}{Lygeros, J.}, \bibinfo{year}{2020}.
\newblock \bibinfo{title}{Cascade control: Data-driven tuning approach based on
  bayesian optimization}.
\newblock \bibinfo{journal}{IFAC-PapersOnLine} \bibinfo{volume}{53},
  \bibinfo{pages}{382--387}.
\newblock \bibinfo{note}{21st IFAC World Congress}.
\bibitem[{Khosravi et~al.(2022)Khosravi, Behrunani, Myszkorowski, Smith,
  Rupenyan and Lygeros}]{Performance_Khosravi2022}
\bibinfo{author}{Khosravi, M.}, \bibinfo{author}{Behrunani, V.N.},
  \bibinfo{author}{Myszkorowski, P.}, \bibinfo{author}{Smith, R.S.},
  \bibinfo{author}{Rupenyan, A.}, \bibinfo{author}{Lygeros, J.},
  \bibinfo{year}{2022}.
\newblock \bibinfo{title}{Performance-driven cascade controller tuning with
  {B}ayesian optimization}.
\newblock \bibinfo{journal}{{IEEE} Transactions on Industrial Electronics}
  \bibinfo{volume}{69}, \bibinfo{pages}{1032--1042}.
\bibitem[{Kim et~al.(2021)Kim, Allmendinger and
  L{\'o}pez-Ib{\'a}{\~{n}}ez}]{Safe_Kim2021}
\bibinfo{author}{Kim, Y.}, \bibinfo{author}{Allmendinger, R.},
  \bibinfo{author}{L{\'o}pez-Ib{\'a}{\~{n}}ez, M.}, \bibinfo{year}{2021}.
\newblock \bibinfo{title}{Safe learning and optimization techniques: Towards a
  survey of the state of the art}, in: \bibinfo{editor}{Heintz, F.},
  \bibinfo{editor}{Milano, M.}, \bibinfo{editor}{O'Sullivan, B.} (Eds.),
  \bibinfo{booktitle}{Trustworthy AI - Integrating Learning, Optimization and
  Reasoning}, \bibinfo{publisher}{Springer International Publishing},
  \bibinfo{address}{Cham}. pp. \bibinfo{pages}{123--139}.
\bibitem[{König et~al.(2023)König, Ozols, Makarova, Balta, Krause and
  Rupenyan}]{Safe_Koenig2023}
\bibinfo{author}{König, C.}, \bibinfo{author}{Ozols, M.},
  \bibinfo{author}{Makarova, A.}, \bibinfo{author}{Balta, E.C.},
  \bibinfo{author}{Krause, A.}, \bibinfo{author}{Rupenyan, A.},
  \bibinfo{year}{2023}.
\newblock \bibinfo{title}{Safe risk-averse {B}ayesian optimization for
  controller tuning}.
\newblock \bibinfo{journal}{IEEE Robotics and Automation Letters} ,
  \bibinfo{pages}{1--8}.
\bibitem[{König et~al.(2021)König, Turchetta, Lygeros, Rupenyan and
  Krause}]{könig2021safe}
\bibinfo{author}{König, C.}, \bibinfo{author}{Turchetta, M.},
  \bibinfo{author}{Lygeros, J.}, \bibinfo{author}{Rupenyan, A.},
  \bibinfo{author}{Krause, A.}, \bibinfo{year}{2021}.
\newblock \bibinfo{title}{Safe and efficient model-free adaptive control via
  {B}ayesian optimization}, in: \bibinfo{booktitle}{2021 IEEE International
  Conference on Robotics and Automation (ICRA)}, pp.
  \bibinfo{pages}{9782--9788}.
\bibitem[{Lee et~al.(2000)Lee, Tan, Huang and Dou}]{lee2000intelligent}
\bibinfo{author}{Lee, T.}, \bibinfo{author}{Tan, K.}, \bibinfo{author}{Huang,
  S.}, \bibinfo{author}{Dou, H.}, \bibinfo{year}{2000}.
\newblock \bibinfo{title}{Intelligent control of precision linear actuators}.
\newblock \bibinfo{journal}{Engineering Applications of Artificial
  Intelligence} \bibinfo{volume}{13}, \bibinfo{pages}{671--684}.
\bibitem[{Mesbah et~al.(2022)Mesbah, Wabersich, Schoellig, Zeilinger, Lucia,
  Badgwell and Paulson}]{mesbah2022fusion}
\bibinfo{author}{Mesbah, A.}, \bibinfo{author}{Wabersich, K.P.},
  \bibinfo{author}{Schoellig, A.P.}, \bibinfo{author}{Zeilinger, M.N.},
  \bibinfo{author}{Lucia, S.}, \bibinfo{author}{Badgwell, T.A.},
  \bibinfo{author}{Paulson, J.A.}, \bibinfo{year}{2022}.
\newblock \bibinfo{title}{Fusion of machine learning and {MPC} under
  uncertainty: {W}hat advances are on the horizon?}, in:
  \bibinfo{booktitle}{2022 American Control Conference (ACC)},
  \bibinfo{organization}{IEEE}. pp. \bibinfo{pages}{342--357}.
\bibitem[{Rasmussen and Williams(2006)}]{gaussian_processes}
\bibinfo{author}{Rasmussen, C.E.}, \bibinfo{author}{Williams, C.K.I.},
  \bibinfo{year}{2006}.
\newblock \bibinfo{title}{Gaussian Processes for Machine Learning}.
\newblock \bibinfo{publisher}{{M}assachusetts {I}nstitute of {T}echnology}.
\bibitem[{Rothfuss et~al.(2022)Rothfuss, König, Rupenyan and
  Krause}]{rothfuss2022metalearning}
\bibinfo{author}{Rothfuss, J.}, \bibinfo{author}{König, C.},
  \bibinfo{author}{Rupenyan, A.}, \bibinfo{author}{Krause, A.},
  \bibinfo{year}{2022}.
\newblock \bibinfo{title}{Meta-learning priors for safe {B}ayesian
  optimization}, in: \bibinfo{booktitle}{Conference on Robot Learning},
  \bibinfo{organization}{PMLR}, \bibinfo{address}{Conference on Robot Learning,
  14-18 December 2022, Auckland, New Zealand}. pp. \bibinfo{pages}{237--265}.
\bibitem[{Skogestad(2023)}]{Advanced_Skogestad2023}
\bibinfo{author}{Skogestad, S.}, \bibinfo{year}{2023}.
\newblock \bibinfo{title}{Advanced control using decomposition and simple
  elements}.
\newblock \bibinfo{journal}{Annual Reviews in Control} \bibinfo{volume}{56},
  \bibinfo{pages}{100903}.
\bibitem[{Srinivas et~al.(2012)Srinivas, Krause, Kakade and
  Seeger}]{Information_Srinivas2012}
\bibinfo{author}{Srinivas, N.}, \bibinfo{author}{Krause, A.},
  \bibinfo{author}{Kakade, S.M.}, \bibinfo{author}{Seeger, M.W.},
  \bibinfo{year}{2012}.
\newblock \bibinfo{title}{Information-theoretic regret bounds for {G}aussian
  process optimization in the bandit setting}.
\newblock \bibinfo{journal}{{IEEE} Transactions on Information Theory}
  \bibinfo{volume}{58}, \bibinfo{pages}{3250--3265}.
\bibitem[{Sui et~al.(2015a)Sui, Gotovos, Burdick and Krause}]{Safe_Sui2015}
\bibinfo{author}{Sui, Y.}, \bibinfo{author}{Gotovos, A.},
  \bibinfo{author}{Burdick, J.}, \bibinfo{author}{Krause, A.},
  \bibinfo{year}{2015}a.
\newblock \bibinfo{title}{Safe exploration for optimization with {G}aussian
  processes}, in: \bibinfo{editor}{Bach, F.}, \bibinfo{editor}{Blei, D.}
  (Eds.), \bibinfo{booktitle}{Proceedings of the 32nd International Conference
  on Machine Learning}, \bibinfo{publisher}{PMLR}, \bibinfo{address}{Lille,
  France}. pp. \bibinfo{pages}{997--1005}.
\bibitem[{Sui et~al.(2015b)Sui, Gotovos, Burdick and Krause}]{safeopt}
\bibinfo{author}{Sui, Y.}, \bibinfo{author}{Gotovos, A.},
  \bibinfo{author}{Burdick, J.W.}, \bibinfo{author}{Krause, A.},
  \bibinfo{year}{2015}b.
\newblock \bibinfo{title}{Safe exploration for optimization with {G}aussian
  processes}, in: \bibinfo{booktitle}{Proceedings of the \nth{32} International
  Conference on Machine Learning}, \bibinfo{address}{Lille, France}. pp.
  \bibinfo{pages}{997--1005}.
\bibitem[{Sukhija et~al.(2023)Sukhija, Turchetta, Lindner, Krause, Trimpe and
  Baumann}]{Scalable_Sukhija2022}
\bibinfo{author}{Sukhija, B.}, \bibinfo{author}{Turchetta, M.},
  \bibinfo{author}{Lindner, D.}, \bibinfo{author}{Krause, A.},
  \bibinfo{author}{Trimpe, S.}, \bibinfo{author}{Baumann, D.},
  \bibinfo{year}{2023}.
\newblock \bibinfo{title}{{GoSafeOpt}: Scalable safe exploration for global
  optimization of dynamical systems}.
\newblock \bibinfo{journal}{Artificial Intelligence} \bibinfo{volume}{320},
  \bibinfo{pages}{103922}.
\bibitem[{Swersky et~al.(2013)Swersky, Snoek and Adams}]{swersky2013multi}
\bibinfo{author}{Swersky, K.}, \bibinfo{author}{Snoek, J.},
  \bibinfo{author}{Adams, R.P.}, \bibinfo{year}{2013}.
\newblock \bibinfo{title}{Multi-task {B}ayesian optimization}, in:
  \bibinfo{editor}{Burges, C.}, \bibinfo{editor}{Bottou, L.},
  \bibinfo{editor}{Welling, M.}, \bibinfo{editor}{Ghahramani, Z.},
  \bibinfo{editor}{Weinberger, K.} (Eds.), \bibinfo{booktitle}{Advances in
  Neural Information Processing Systems ({NIPS 2013})},
  \bibinfo{publisher}{Curran Associates, Inc.}. pp.
  \bibinfo{pages}{2004--2012}.
\bibitem[{Xu and Saleh(2021)}]{xu2021machine}
\bibinfo{author}{Xu, Z.}, \bibinfo{author}{Saleh, J.H.}, \bibinfo{year}{2021}.
\newblock \bibinfo{title}{Machine learning for reliability engineering and
  safety applications: Review of current status and future opportunities}.
\newblock \bibinfo{journal}{Reliability Engineering \& System Safety}
  \bibinfo{volume}{211}, \bibinfo{pages}{107530}.
\bibitem[{Zagorowska et~al.(2023)Zagorowska, Balta, Behrunani, Rupenyan and
  Lygeros}]{zagorowska2022efficient}
\bibinfo{author}{Zagorowska, M.}, \bibinfo{author}{Balta, E.C.},
  \bibinfo{author}{Behrunani, V.}, \bibinfo{author}{Rupenyan, A.},
  \bibinfo{author}{Lygeros, J.}, \bibinfo{year}{2023}.
\newblock \bibinfo{title}{Efficient sample selection for safe learning}, in:
  \bibinfo{booktitle}{{IFAC World Congress 2023}}.
\newblock \bibinfo{note}{Online:
  \href{https://www.research-collection.ethz.ch/handle/20.500.11850/615589}{20.500.11850/615589}}.

\end{thebibliography}

  \end{document}